\documentclass[amsmath,12pt,amssymb,preprint,prd,aps,nofootinbib]{revtex4}
\usepackage{amsfonts} %\usepackage{citesort}
\usepackage{graphicx} % Include figure files
\usepackage{epsfig}
\usepackage{multirow}
\usepackage{xcolor}
\usepackage{bm}% bold math

\newcommand{\diracslash}[1]{#1\llap{/\kern2pt}}

\newcommand{\be}{\begin{equation}}
\newcommand{\ee}{\end{equation}}
\newcommand{\bea}{\begin{eqnarray}}
\newcommand{\eea}{\end{eqnarray}}
\newcommand{\ba}[1]{\begin{array}{#1}}
\newcommand{\ea}{\end{array}}

\newcommand{\bt}{\begin{tabular}}
\newcommand{\et}{\end{tabular}}

\newcommand{\beas}{\begin{eqnarray*}}
\newcommand{\eeas}{\end{eqnarray*}}

\begin{document}
\title{Open Heavy flavor mesons in hot asymmetric\\
strange hadronic matter
-- a QCD sum rule approach} 
\author{Amruta Mishra}
\email{amruta@physics.iitd.ac.in}
\affiliation{Department of Physics, Indian Institute of Technology Delhi, 
Hauz Khas, New Delhi -- 110016, India}
\vskip -1.5in
\begin{abstract}
The in-medium masses of the pseudoscalar open charm 
($D$, $\bar D$, $D_s$ and $\bar {D_s}$) 
and open bottom 
($B$, $\bar B$, $B_s$ and $\bar {B_s}$) 
mesons in hot asymmetric strange hadronic matter 
are studied within a QCD sum rule approach. 
These are computed using the medium modifications 
of the light quark condensates ($\langle{\bar{q_i}}{q_i}\rangle$,
with $(q_i,i=1,2,3)\equiv(u,d,s)$), the scalar
gluon condensate, $\left\langle \frac{\alpha_{s}}{\pi} 
G^a_{\mu\nu} {G^a}^{\mu\nu} \right\rangle$,
the twist-2 tensorial gluon operator, 
$\left\langle \frac{\alpha_{s}}{\pi} 
\Big (G^a_{\mu\sigma} {{G^a}_\nu}^{\sigma} u^\mu u^\nu
-\frac{1}{4} G^a_{\mu\sigma} {G^a}^{\mu \sigma}\Big)
\right\rangle$ (where $u^\mu$ is the 4-velocity of the medium)
and other operators in the operator product expansion
upto mass dimension 5. Within a chiral SU(3) model, the quark
condensates are obtained from the medium modifications of the 
non-strange and strange scalar-isoscalar fields ($\sigma$ and $\zeta$), 
and the scalar-isovector field, $\delta$, whereas, the gluon 
(scalar and twist-2) condensates are computed from the in-medium 
value of the dilaton field, $\chi$, which is incorporated 
within the chiral model to mimic the broken scale invariance of QCD.
The mixed quark-gluon condensate 
$\langle \bar {q_i} g_s \sigma . G q_i \rangle$
is calculated from the in-medium 
quark condensate $\langle \bar {q_i} q_i \rangle$.
The splittings of the masses of
the $D-\bar D$ ($B-\bar B)$, as well as $D_s-\bar {D_s}  
(B_s-\bar {B_s})$ in the hadronic medium are due to
the odd part of the spectral function.
For the strange-charm (strange-bottom) mesons ($(q_i,i=3)\equiv s$), 
the value of $\langle {q_i}^\dagger {q_i} \rangle$
dominates over the contributions from the other operators,
$\langle {q_i}^\dagger iD_0^2 {q_i}\rangle$ 
and $\langle {q_i}^\dagger g_s \sigma . G {q_i} \rangle$,
of the odd part of the sectral function.
This is observed as opposite behaviour of the particle-antiparticle
mass splittings with the mass of the particle to be larger (smaller) 
than the antiparticle mass in nuclear (hyperonic) matter. 
The density and isospin asymmetry effects are observed
to be the dominant medium effects which might have observable 
consequences on the production of charmonia and open charm mesons, 
and, on the production ratios, $D^+/D^-$ and $D^0/\bar {D^0}$, 
in asymmetric heavy ion collisions in the CBM experiment at FAIR
at the future facility at GSI.
 
\end{abstract}
\maketitle
\newpage

\section{Introduction}
\label{intro}
The study of the properties of hadrons under extreme conditions
of temperature and/or density is an important subject of research
in strong interaction physics. The topic is of direct relevance 
for the experimental observables of the relativistic 
heavy-ion collision experiments, as well as, is relevant for
study of the bulk properties of compact objects, e.g., the
neutron stars. 
The light vector mesons have been extensively studied in the literature,
as the medium modifications of these mesons can be observed
in the dilepton spectra arising from heavy ion collision experiments
\cite{Rapp_Wambach}.
In the recent years, a lot of studies on the
in-medium properties of heavy flavor hadrons have been initiated
due to its relevance and possible observability 
in the relativistic heavy ion collision experiments 
\cite{Prog_Nucl_Part_Phys_2017_Hosaka_et_al}.
The spectral modification of the charm hadrons 
will be investigated via $\bar p$-A collisions at FAIR 
at the future facility at GSI, Darmstadt 
\cite{Yamagata-Sekihara_et_al,Rapp}
and charm production using pion beams on proton in
Japan Proton Accelerator Research Complex (J-PARC)
\cite{Kim1,Kim2,Garzon}.
The charm (bottom) meson-nucleon interaction can also lead to possibility
of formation of charm (bottom) mesons bound in atomic nuclei.

The properties of heavy flavor mesons in the (hot) hadronic matter
have been studied extensively in the literature.
These have been investigated using the potential models 
\cite{potential_models},
%\cite{eichten_1,eichten_2,satz_1,satz_2,satz_3,satz_4,satz_5,repko,
%Ebert,Bonati_pot_model,Yoshida_Suzuki_heavy_flavour_meson_strong_B},
the coupled channel approach
\cite{coupled_channel_approach},
%\cite{ltolos,ljhs,mizutani_1,mizutani_2,HL,tolos_heavy_mesons_1,tolos_heavy_mesons_2}, 
the quark meson coupling (QMC) model
\cite{QMC_models,krein_jpsi,N_Quarkonia_krein}
heavy quark symmetry and interaction
of these mesons with nucleons via pion exchange \cite{Yasui_Sudoh_pion},
heavy meson effective theory
\cite{Yasui_Sudoh_heavy_meson_Eff_th}, studying the heavy flavor meson as
an impurity in nuclear matter \cite{Yasui_Sudoh_heavy_particle_impurity},
and the QCD sum rule approach
\cite{open_heavy_flavour_qsr_1,kimlee,klingl,amarvjpsi_qsr,moritalee_1,moritalee_2,moritalee_3,moritalee_4,leemorita2,open_heavy_flavour_qsr_2,Hilger_thesis,open_heavy_flavour_qsr_3,open_heavy_flavour_qsr_4,Wang_heavy_mesons_1,Wang_heavy_mesons_2,arvind_heavy_mesons_QSR_1,arvind_heavy_mesons_QSR_2,arvind_heavy_mesons_QSR_3}.
Using leading order QCD formula \cite{pes1,pes2,voloshin},
the medium modifications of the masses of the charmonium states 
due to the medium change of the scalar gluon condensate
were studied in a linear density approximation in Ref.\cite{leeko},
which showed much larger mass shifts for the excited states
($\psi(3686)$ and $\psi(3770)$), as compared to the mass shift of $J/\psi$.
Within a chiral effective model \cite{sche1,heide1,paper3,kristof1,hartree},
the in-medium masses of the charmonium (bottomonium) states 
\cite{amarvdmesonTprc,amarvepja,AM_DP_upsilon}
as well as of the open heavy flavor mesons
\cite{amdmeson,amarindamprc,amarvdmesonTprc,amarvepja,DP_AM_Ds,DP_AM_bbar,DP_AM_Bs}
have been studied. Within the chiral effective model, 
the mass modifications of the quarkonium states arise
due to the medium modification of a scalar dilaton field,
which is introduced in the model to mimic the broken scale
invariance of QCD. The open heavy flavor mesons 
have been studied due to their interactions with the baryons
and the scalar mesons in the hadronic medium.
The model has been used to study the partial decay widths
of the heavy quarkonium states to the open heavy flavor mesons,
in the (magnetized) hadronic medium
\cite{amarvepja,charmdecay_mag} 
using a light quark pair creation model \cite{friman},
namely the $^3P_0$ model \cite{3p0_1,3p0_2,3p0_3,3p0_4} as well as
using a field theoretical model for composite hadrons
\cite{amspmwg,amspm_upsilon,charmdw_mag,upslndw_mag}.
Within the chiral effective model, in the presence of an external
magnetic field, the effects of magnetized Dirac sea have been investigated
on the properties of heavy flavor mesons in the hadronic medium 
\cite{HQ_DW_DS_AM_SPM_2023,Charmonium_DS,Open_Bottom_DS}.
The magnetized Dirac sea, accounting for
the anomalous magnetic moment of nucleons is observed to lead
to the phenomenon of inverse magnetic catalysis at finite densities for
symmetric as well as asymmetric nuclear matter in presence of strong
magnetic fields. Recently, the effects of the Dirac sea as well as mixing
of the pseudoscalar and vector mesons on the
masses, decay widths and the production cross-sections
of the charmonium states in hot magnetized strange hadronic matter
have been investigated in Ref. \cite{AMAKSPM24}.

Within the chiral SU(3) model, using the medium changes 
of the light quark condensates 
($\langle \bar u u \rangle$, $\langle \bar d d \rangle$ and
$\langle \bar s s \rangle$), obtained from the non-strange
isoscalar ($\sigma$) isovector ($\delta$) and strange
($\zeta$) scalar fields
and gluon condensates calculated from the scalar dilaton field,
$\chi$, the light vector mesons ($\omega$, $\rho$ and $\phi$),
in (magnetized) hadronic matter have been
studied within the framework of QCD sum rule approach
 \cite{am_vecmeson_qsr,vecqsr_mag}.
The effects of the magnetized Dirac sea
on the spectral properties of light vector and axial-vector mesons 
in hot magnetized nuclear matter \cite{Spectr_rho_A1_DS}
and of the ground state heavy quarkonia
\cite{QCDSR_Heavy_Quarkonia} have also been studied 
with the QCD sum rule approach.
In the present work, we investigate the masses 
of the open charm ($D$, $\bar D$, $D_s$)
and open bottom ($B$, $\bar B$, $B_s$) mesons
in isospin asymmetric strange hadronic matter
at finite temperature. 
In section II, we briefly describe the QCD sum rule approach 
used to compute the masses of the open charm and open bottom
mesons, retaining the operators upto dimension 5 in the operator
product expansion of the spectral function.
Unlike the case of vacuum, in the hadronic medium, 
there is splitting between the masses 
of the particles and their antiparticles arising due to
the odd part of the spectral function.
In section III, we describe the chiral SU(3) model
used to compute the in-medium values of the quark (light and
strange), the gluon (scalar and twist-2) and the mixed quark-gluon 
operators, which are used for the calculation of the in-medium
masses of the pseudoscalar open charm and open bottom mesons.
Section IV discusses the results and in section V, we summarize
the findings of the present work. 

\section{QCD sum rule approach}
The QCD sum rule approach relates the hadronic properties,
encoded in spectral functions, to the QCD condensates 
\cite{Cohen_Prog_Part_Nucl_Phys,QCD_Cond_Gubler_Satow_Prog_Part_Nucl_Phys}.
The basic ingredient is the two-point correlation function
given as
\begin{equation}
\Pi (q)=i \int d^4 x e^{-iq.x} \langle \Omega| T(j(x)j(0)^\dagger)
|\Omega \rangle,
\end{equation}
where, $|\Omega\rangle$ is the physical ground state.
Unlike the case of vacuum, the two-point correlation function
in the hadronic medium has contributions which are odd in $q_0$,
and, can be written in terms of even and odd functions 
of $q_0$ as
\begin{equation}
\Pi (q_0,\vec q)=\Pi^e(q_0,\vec q)+q_0\Pi^o (q_0,\vec q),
\end{equation}
with $\Pi^e(q_0,\vec q)=\frac{1}{2}(\Pi(q_0,\vec q)
+\Pi(-q_0,\vec q))$ and 
$\Pi^o(q_0,\vec q)=\frac{1}{2q_0}(\Pi(q_0,\vec q)
-\Pi(-q_0,\vec q))$.
It might be noted here that $\Pi^{e(o)}(q_0,\vec q)$ depends on $q_0^2$.
The Borel transformed sum rules for the even and odd parts,
for $\vec q=0$, are given as
\cite{open_heavy_flavour_qsr_2}
\begin{eqnarray}
{\cal B}[\Pi^e_{OPE}(\omega^2,\vec q=0)](M^2)=
\frac{1}{\pi} 
\Big [
\int_{s_0^-}^ {s_0^+}
{\rm d}s\; {\rm {Im} }\Pi^{(ph)}(s) 
 + \Big (\int_{-\infty}^{s_0^-} 
+\int_{s_0^+}^\infty \Big) {\rm d}s\; {\rm {Im} }\Pi (s)
\Big] s \; e^{-\frac{s^2}{M^2}},
\label{Pi_e_M2}
\end{eqnarray}
and
\begin{eqnarray}
 {\cal B}[\Pi^o_{OPE}(\omega^2,\vec q=0)](M^2)
= \frac{1}{\pi} \Big [
\int_{s_0^-}^ {s_0^+}
{\rm d}s\; {\rm {Im} }\Pi^{(ph)}(s)
+\Big (\int_{-\infty}^{s_0^-} 
+\int_{s_0^+}^\infty \Big) {\rm d}s\; {\rm {Im} }\Pi(s)
\Big] e^{-\frac{s^2}{M^2}},
\label{Pi_o_M2}
\end{eqnarray}
In the above equations, $\omega^2=-q_0^2$, 
the subscript `$OPE$' denotes the operator product expansion 
$\langle \Omega| T(j(x)j(y)^\dagger) |\Omega \rangle
=\sum _n C_n (x-y) \langle \Omega | O_n |\Omega \rangle$, 
where $\langle \Omega | O_n |\Omega \rangle$ are the QCD condensates
for the operators $O_n$, $C_n(x-y)$ are the Wilson 
coefficients, $s_0^+ (s_0^-)$ is the continuum threshold
for the particle (antiparticle) and $M$ is the Borel mass.
The open charm and open bottom mesons have been studied
in symmetric nuclear matter within the QCD sum rue approach 
keeping operators upto dimension 5 in
the OPE side of the spectral function 
in Ref.  \cite{open_heavy_flavour_qsr_2}.
In the present work, we study the in-medium masses 
of these pseudoscalar mesons
in hot isospin asymmetric strange hadronic matter, again,
retaining QCD operators upto dimension 5 in the operator
product expansion.
The spectral functions of the open heavy flavor mesons
are obtained using the current operators for the particles ($p_i$),
$j_i=j^p_i \equiv i {\bar {q_i}} \gamma_5 Q$, 
with $(q_i, i=1,2,3) \equiv (u,d,s)$, 
where, the heavy quark $Q=c$ corresponds to $D^0$, 
$D^+$ and $D_s^+$ mesons, and,
$Q=b$ corresponds to $B^-$, $\bar {B^0}$ and 
$\bar {B_s}^0$ mesons respectively.
The current operator is $j_i^\dagger \equiv  i {\bar Q} \gamma_5 {q_i}$,
for the antiparticles ($\bar {D^0}$, $D^-$, $D_s^-$
for charm sector and $B^+$, $B^0$ and ${B_s}^0$
for the bottom sector). 
For the phenomenological hadronic spectral function,
we take the ansatz \cite{open_heavy_flavour_qsr_2}
\begin{equation} 
{\rm Im}\Pi^{(ph)}(s)
=\pi F_+ \delta (s-m_+) -\pi F_- \delta (s+m_-),
\label{spectral_s}
\end{equation}
where $m_\pm$ corresponds to the mass of the particle (antiparticle).
In equations (\ref{Pi_e_M2}) and (\ref{Pi_o_M2}),
we change the integration variable from $s$ to $\tilde s\;(=s^2)$ 
in the integrals corresponding to the continuum parts
of the spectral functions, and,
choose the continuum thresholds for the particle and antiparticle,
to be $s_0^\pm=\pm \sqrt{{\tilde s}_0}$, similar to the case of vacuum.
We also assume, according to semi-local duality hypothesis, 
that ${\rm Im}\Pi ({\tilde s})$ corresponding 
to the particle and antiparticle 
continuum parts of the spectral function
%($-\infty \lt s \lt s_0^-$ and $s_0^+ \lt s \lt \infty$) 
on the right hand side of the equations 
(\ref{Pi_e_M2}) and (\ref{Pi_o_M2})
to be same as the perturbative contribution of the
spectral function, ${\rm Im}\Pi_{OPE}^{per}(s)$, 
as calculated in operator product expansion \cite{Hilger_thesis}.
It might be noted here that for the odd part of the 
spectral function, the perturbative contributions 
from the particle and antiparticle cancel with each other
for the choice of the continuum thresholds 
$s_0^\pm=\pm \sqrt{{\tilde s}_0}$, as in the present work.
The Borel transformed sum rules for the even and odd parts of the 
spectral function are then obtained as 
\cite{open_heavy_flavour_qsr_2}
\begin{eqnarray}
&& {\cal B}[\Pi^e_{OPE}(\omega^2,\vec q=0)](M^2)=
\frac{1}{\pi}\int _{m_Q^2}^{\infty} {\rm d}{\tilde s} 
{\rm Im}\; \Pi^{per}_{OPE} ({\tilde s},\vec q=0)
{e^{-{\tilde s}/M^2}} 
+ e^{-{m_Q^2}/{M^2}} P^e (M^2) \nonumber \\
&=& m_+F_+e^{-{m_+^2}/{M^2}}+m_-F_-e^{-{m_-^2}/{M^2}}
+\frac{1}{\pi}\int _{{\tilde s}_0}^\infty  {\rm d}{\tilde s}
{\rm Im}\Pi^{per}_{OPE}({\tilde s})
e^{-{\tilde s}/{M^2}} 
\label{spectr_even_M^2}
\end{eqnarray}
with
\begin{eqnarray} 
P^e (M^2)
&=& -m_Q\langle {\bar {q_i}} {q_i} \rangle
+\frac{1}{2} m_{q_i}\langle {\bar {q_i}}{q_i}\rangle
\Big (1+\frac{m_Q^2}{M^2}\Big)
+\frac{m_Q}{2M^2}\Big(\frac{m_Q^2}{2M^2}-1\Big)
%%\langle \bar {q_i} g_s \sigma_{\mu \nu} G^{\mu \nu} q_i\rangle
\langle \bar {q_i} g_s \sigma \cdot G q_i\rangle \nonumber \\
&+&\frac{1}{12}\langle \frac{\alpha_s}{\pi} G_{\mu \nu} 
G^{\mu \nu}\rangle +
\Bigg [ \Big ( \frac{7}{18} +\frac{1}{3} 
\ln \Big(\frac{\mu ^2 m_Q^2}{M^4}\Big) -\frac{2 \gamma_E}{3}\Big)
\Big(\frac{m_Q^2}{M^2}-1\Big)-\frac{2}{3}\frac{m_Q^2}{M^2}\Bigg]
\nonumber \\
&\times &\langle \frac{\alpha_{s}}{\pi} 
\Big (G^a_{\mu\sigma} {{G^a}_\nu}^{\sigma} u^\mu u^\nu
-\frac{1}{4} G^a_{\mu\sigma} {G^a}^{\mu \sigma}\Big)
\rangle 
+ 2 \Big (\frac{m_Q^2}{M^2} -1\Big) 
\langle {q_i}^\dagger i D_0 q_i \rangle 
\nonumber \\
&+& 4 \Big ( \frac{m_Q^3}{2 M^4}-\frac{m_Q}{M^2}
\Big) \Big[ \langle \bar {q_i}D_0^2 {q_i}\rangle
-\frac{1}{8} 
\langle \bar {q_i} g_s \sigma \cdot G q_i\rangle \Big],
\label{pe}
\end{eqnarray}
and,
\begin{eqnarray}
{\cal B}[\Pi^o_{OPE}(\omega^2,\vec q=0)](M^2)
= e^{-{m_Q^2}/{M^2}}  P^o (M^2)
= F_+ e^{-{m_+^2}/{M^2}}-F_- e^{-{m_-^2}/{M^2}}
\label{spectr_odd_M^2}
\end{eqnarray}
with
\begin{eqnarray} 
P^o (M^2)=\langle {q_i}^\dagger {q_i}\rangle 
-4 \Big (\frac{m_Q^2}{2M^4}-\frac{1}{M^2}\Big ) 
\langle {q_i}^\dagger D_0^2 {q_i}\rangle
-\frac{1}{M^2} 
\langle {q_i}^\dagger g_s \sigma \cdot G q_i\rangle. 
\label{po}
\end{eqnarray}
The QCD condensates corresponding to the even and odd parts of the 
spectral function occur in the polynomials $P^{e} (M^2)$ and 
$P^{o} (M^2)$, given by equations (\ref{pe}) and (\ref{po})
respectively. For the polynomial $P^e (M^2)$, 
the first and second terms of the 
arise due to the quark-antiquark condensate 
($\langle {\bar {q_i}} q_i\rangle$), 
retaining the term proportional to the light quark mass ($m_{q_i}$) 
\cite{open_heavy_flavour_qsr_1} as given by the second term.
The first term, due to the large mass of the heavy quark ($m_Q$) 
is the dominant contribution to the mass of the open charm (bottom)
meson, arising from the dimension 3 quark-antiquark condensate
in the operator product expansion.
The strong coupling constant, $\alpha_s$ is given as
$\alpha_s=4\pi /\Big [\big( (11-2 N_f/3)
\rm {ln} (\mu ^2/\Lambda_{QCD}^2)\big)\Big]$, with 
$\Lambda_{QCD}=500$ MeV, and, $N_f$=4 (5) for the
study of the open charm (bottom) mesons.
$\mu$, the renormalization scale, is taken to be mass of the
largest quark in the system, i.e, the mass of the
charm (bottom) quark, for investigating the mass of the 
open charm (bottom) mesons \cite{open_heavy_flavour_qsr_2},
$\gamma_E$ is the Euler constant. The other terms 
are the higher mass dimension operators, which include the 
dimension-4  scalar and twist-2 gluon condensates,
the mixed quark-gluon operator (the abbreviation
$\sigma \cdot G$ is used for $\sigma_{\mu \nu} G^{\mu \nu}$), 
and terms proportional
to the operators $\langle {q_i}^\dagger i D_0 q_i \rangle $
and $\langle \bar {q_i}D_0^2 {q_i}\rangle$.
In the term proportional to the twist-2 gluon condensate
operator, $u^\mu$ is the 4-velocity of the hadronic medium.
We study the spectral functions of the open heavy flavor mesons
in the rest frame of the hadronic medium, for which
$u^\mu=(1, \vec u=0$).
In the odd part of the Borel transformed spectral function,
the dominant contribution in equation (\ref{po}) is from the
number density of the quark ($\langle {q_i}^\dagger {q_i}\rangle$),
along with the contributions from the dimension 5 operators,  
$\langle \bar {q_i}D_0^2 {q_i}\rangle$ and
$\langle {q_i}^\dagger g_s \sigma \cdot G q_i\rangle$. 
The masses for the particles and antiparticle, $m_\pm$,
corresponding to the open heavy flavor mesons,
are solved by using the spectral function 
given by equation (\ref{spectral_s})
in the equations (\ref{Pi_e_M2}) and  (\ref{Pi_o_M2})
\cite{open_heavy_flavour_qsr_2}.
The expressions for the masses of the particle
and antiparticle are given as \cite{Hilger_thesis}
\begin{equation}
m_\pm=\Bigg [\frac{1}{4} \Big(\frac{gf'-fg'}{f^2+gg'}\Big)^2
-\Big (\frac{ff'+{g'}^2}{f^2+gg'}\Big)\Bigg]^{1/2}
\pm \frac{1}{2}\Big( \frac{gf'-fg'}{f^2+gg'}\Big)
\label{mass_pm}
\end{equation}
where, $f=\frac{1}{\pi}\int _{m_Q^2}^{{\tilde s}_0} {\rm d}{\tilde s} 
{\rm Im}\; \Pi_{OPE}^{per} ({\tilde s},\vec q=0) 
e^{-{\tilde s}/M^2} 
+ e^{-{m_Q^2}/{M^2}} P^e (M^2)$, with $P^e(M^2)$
given by equation (\ref{pe}), $g=e^{-{m_Q^2}/{M^2}} P^o (M^2)$,
the odd part of the spectral function as 
given by equation (\ref{spectr_odd_M^2}), 
and, $f'$ and $g'$ denote the derivatives 
of $f$ and $g$ with respect to $1/{M^2}$.
As can be observed from  eqn (\ref{mass_pm}), 
the mass difference between the particles 
($D^0$, $D^+$ and $D_s^+$) and their antiparticles 
($\bar {D^0}$, ${D^-}$ 
and $D_s^-$) for the charm sector, and between the particles 
($B^-$, $\bar {B^0}$ and ${\bar {B_s}}^0$) 
and their antiparticles ($B^+$, $B^0$ and 
$B_s^0$) for the bottom sector,
in the hadronic medium arise from the values of
$\langle {q_i}^\dagger {q_i} \rangle$, 
$\langle {q_i}^\dagger iD_0^2 {q_i}\rangle$
and $\langle {q_i}^\dagger g_s \sigma . G {q_i} \rangle$
of the odd part of the spectral function ($g$), hence due to
the expectation values of the QCD operators
$\langle {q_i}^\dagger {q_i}\rangle$, 
$\langle {q_i}^\dagger D_0^2 {q_i}\rangle$ and
$\langle {q_i}^\dagger g_s \sigma \cdot G q_i\rangle$ 
in $P^o(M^2)$, given by equation (\ref{po}).
The in-medium QCD condensates are 
calculated within a chiral SU(3) model, as described in the
following section.

\section{Chiral SU(3) model}

In this section, we briefly describe the chiral SU(3) model 
\cite{paper3,hartree,kristof1,AMAKSPM24} which is used to obtain
the in-medium QCD condenstaes in the hot asymmetric strange hadronic
matter. The expectation values of these QCD operators
are central to the calculation of the in-medium masses 
of the open charm and open bottom mesons within the 
framework of the QCD sum rule approach, as described 
in the previous section.
The chiral SU(3) model \cite{paper3,hartree,kristof1}
is based on a non-linear realization of chiral symmetry 
\cite{Weinberg,coleman,Bardeen} 
and incorporates the broken scale invariance
of QCD through the introduction of a scalar dilaton field, $\chi$
\cite{sche1,heide1}.
The Lagrangian density of chiral SU(3) model is written as \cite{paper3}
\begin{equation}
{\cal L} = {\cal L}_{kin}+\sum_{W=X,Y,V,A,u} {\cal L}_{BW} + 
{\cal L}_{vec} + {\cal L}_{0} +
{\cal L}_\chi +
% {\cal L}_{scale\;break}+
{\cal L}_{SB}.
\label{genlag}
\end{equation}
In Eq.(\ref{genlag}), ${\cal L}_{kin}$ is the kinetic energy term, 
${\cal L}_{BW}$ is the baryon-meson interaction term, where, 
$W=X,Y,V,A,u$ correspond to the interactions of the baryons with
the scalar, pseudoscalar singlet, vector meson, axial-vector meson 
and pseudoscalar octet mesons respectively.
The baryons-spin-0 meson interaction term generates the baryon masses.
${\cal L}_{vec}$  describes 
the dynamical mass generation of the vector mesons via couplings to the 
scalar mesons and contain additionally quartic self-interactions of the 
vector fields. ${\cal L}_{0}$ contains the meson-meson interaction terms, 
${\cal L}_{\chi}$ is the QCD scale breaking term in the Lagrangian density, 
including a logarthimic potential of the scalar dilaton field, $\chi$,
and, ${\cal L}_{SB}$ describes the explicit chiral symmetry breaking.
In the present study of isospin asymmetric strange hadronic matter
at finite temperature,
the mean field approximation is used, in which
the meson fields are replaced by their expectation values.
The meson fields which have non-zero expectation values 
are the scalar ($\sigma, \zeta, \delta)$
and the vector fields, $\omega^\mu \rightarrow \omega \delta^{\mu 0}$,
$\rho^{\mu a}\rightarrow \rho \delta^{\mu 0} \delta^{a 3}$
and, $\phi^\mu \rightarrow \phi \delta^{\mu 0}$, and, the expectation
values of the other mesons are zero.
We also use the approximations 
$\bar \psi_i \psi_j \rightarrow \delta_{ij} \langle \bar \psi_i \psi_i
\rangle \equiv \delta_{ij} \rho_i^s $
and $\bar \psi_i \gamma^\mu \psi_j \rightarrow \delta_{ij} \delta^{\mu 0} 
\langle \bar \psi_i \gamma^ 0 \psi_i
\rangle \equiv \delta_{ij} \delta^{\mu 0} \rho_i $, 
where, $\rho_i^s$ and $\rho_i$ are the scalar and number density 
of baryon of species, $i$. 
In the above approximation, the terms which contribute 
to the baryon-meson interaction,
${\cal L}_{BW}$ of equation (\ref{genlag}) 
arise from the interactions due to the scalar ($S$)
and the vector ($V$) mesons, and, is given as
\begin{equation}
{\cal L}_{BS}+ {\cal L}_{BV}=\sum _i {\bar  \psi}^i
(g_{\sigma i} \sigma+g_{\zeta i} \zeta+g_{\delta i} \delta
-g_{\omega i}\gamma^0\omega -g_{\rho i}\gamma^0\rho
 -g_{\phi i}\gamma^0\phi)\psi^i. 
\label{BS_BV}
\end{equation}
The explicit symmetry breaking term in the chiral model 
is given by
\begin{eqnarray}
{\cal L} _{SB} & = & 
-\left( \frac{\chi}{\chi_{0}}\right) ^{2} 
{\rm Tr} \left [ {\rm diag} \left (
\frac{m_{\pi}^{2} f_{\pi}}{2} (\sigma+\delta), 
\frac{m_{\pi}^{2} f_{\pi}}{2} (\sigma-\delta), 
\Big( \sqrt{2} m_{K}^{2}f_{K} - \frac{1}{\sqrt{2}} 
m_{\pi}^{2} f_{\pi} \Big) \zeta \right) \right ],
\label{lag_SB_MFT}
\end{eqnarray}
which reduces to eqn. (8), and in QCD, 
this term is given as ${\cal L}_{SB}^{QCD}
=-Tr[{\rm diag}(m_u \bar u u,m_d \bar d d, m_s \bar s s)]$.
In the above, the matrices, whose traces give the 
corresponding Lagrangian densities
have been explicitly written down, and a comparison of their 
matrix elements relates the quark condensates 
to the values of the scalar fields 
as \cite{am_vecmeson_qsr}
\begin{eqnarray}
m_u\langle \bar u u \rangle 
%=left( \frac{\chi}{\chi_{0}}\right) ^{2} 
& = &\Big (\frac{\chi}{\chi_{0}}\Big) ^{2} 
\frac{m_{\pi}^{2} f_{\pi}}{2} (\sigma+\delta),\;
%\nonumber\\
%\label{nsubu}
m_d \langle \bar d d \rangle
=\left( \frac{\chi}{\chi_{0}}\right) ^{2} 
\frac{m_{\pi}^{2} f_{\pi}}{2} (\sigma-\delta),
\nonumber \\
%\label{nsdbd}
m_s\langle \bar s s \rangle 
& =& \left( \frac{\chi}{\chi_{0}}\right) ^{2} 
\Big( \sqrt {2} m_{K}^{2}f_{K} - \frac {1}{\sqrt {2}} 
m_{\pi}^{2} f_{\pi} \Big) \zeta.
%\label{sbs}
\label{qqbar}
\end{eqnarray}
The medium modifications of the quark condensates 
$\langle \bar { q_i} q_i\rangle$, with $q_i\equiv u,d,s$ for $i=1,2,3$,
are obtained from the medium modifications of the scalar fields
($\sigma$ and $\zeta$ and $\delta$).

\begin{figure}
\vskip -2.in
\includegraphics[width=14cm,height=16cm]{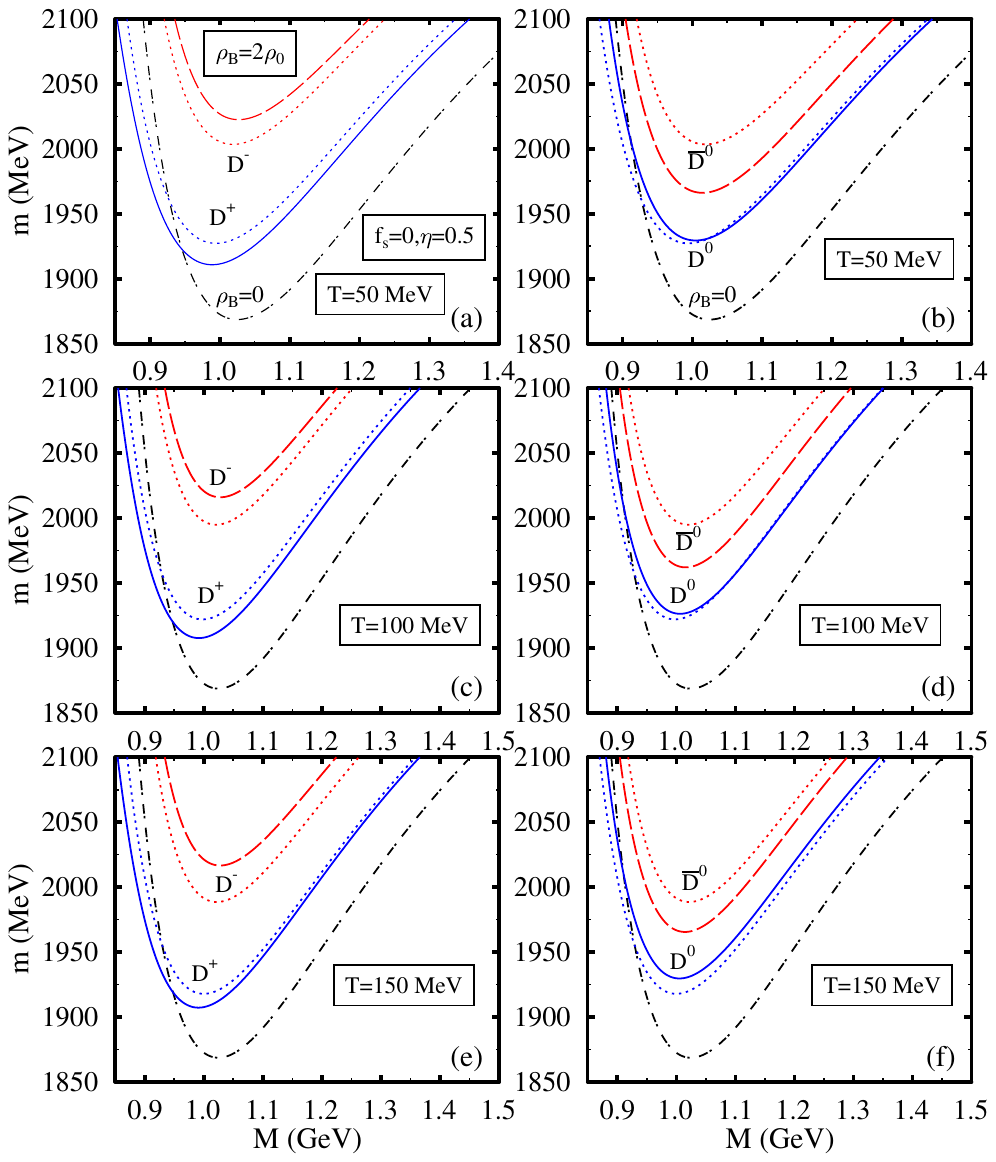} 
\vskip -0.3in
 \caption{
Masses of $D^+ (D^-)$ (in panels (a), (c) and (e)) 
and $D^{0}(\bar {D^0})$ (in panels (b), (d) and (f)) 
mesons are plotted as functions of the Borel mass M 
for T=50, 100 and 150 MeV in nuclear matter ($f_s$=0)
for baryon density $\rho_B=2\rho_0$ and $\eta=0.5$,
as the solid (dashed) lines. These are
compared to the case of $\eta$=0 (shown as dotted lines). 
The Borel curve for $\rho_B=0$ is also shown in the figure
as the dot-dashed line.}
\label{amddbarfs0_2rhb0}
 \end{figure}
\begin{figure}
\vskip -1.1in
\includegraphics[width=14cm,height=16cm]{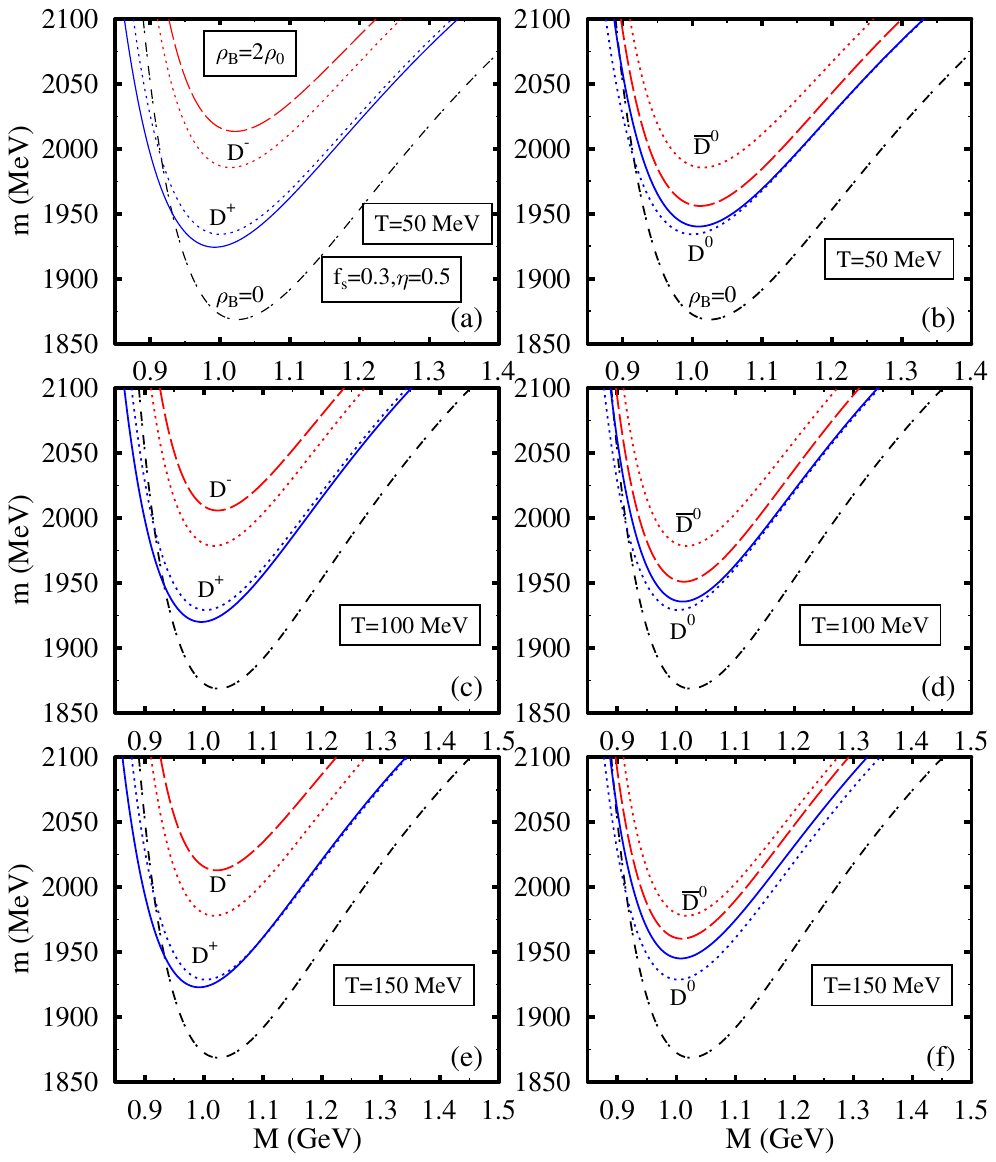} 
\vskip -0.3in
 \caption{
%Same as Fig. \ref{amddbarfs0_2rhb0}, with $f_s$=0.3.
Masses of $D^+ (D^-)$ (in panels (a), (c) and (e)) 
and $D^{0}(\bar {D^0})$ (in panels (b), (d) and (f)) 
mesons are plotted as functions of the Borel mass M 
for T=50, 100 and 150 MeV in hyperonic matter (with $f_s$=0.3)
for baryon density $\rho_B=2\rho_0$ and $\eta=0.5$,
as the solid (dashed) lines. These are
compared to the case of $\eta$=0 (shown as dotted lines). 
The Borel curve for $\rho_B=0$ is also shown in the figure
as the dot-dashed line.
}
\label{amddbarfs3_2rhb0}
 \end{figure}

\begin{figure}
\vskip -2.in
\includegraphics[width=14cm,height=16cm]{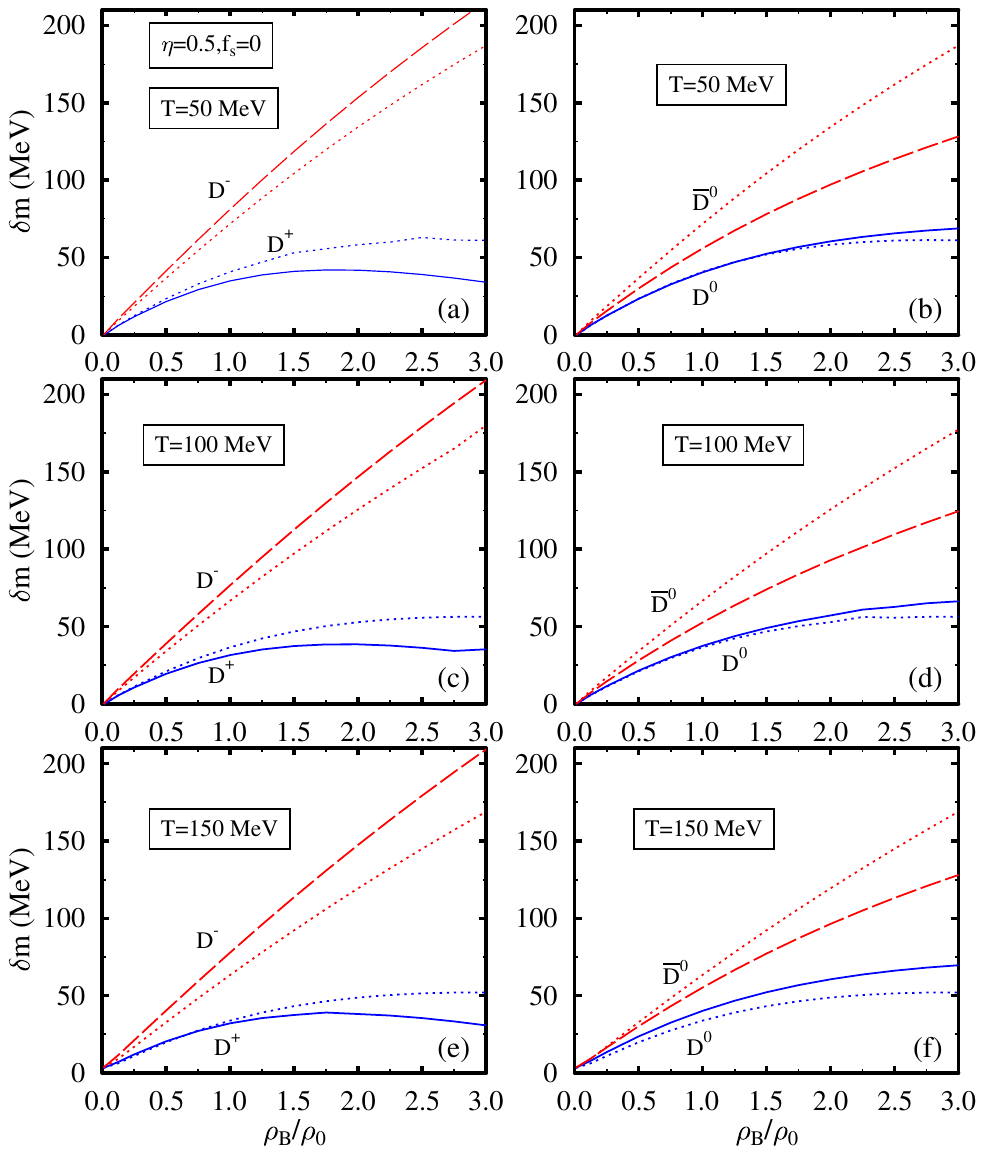} 
\vskip -0.3in
 \caption{
Mass shifts of $D^+(D^-)$ (in panels (a), (c) and (e)) 
and $D^{0}(\bar {D^0})$ (in panels (b), (d) and (f)) 
mesons are plotted as functions of the baryon density
in units of nuclear saturation  density
for T=50, 100 and 150 MeV in nuclear matter ($f_s$=0)
for $\eta=0.5$, as the solid (dashed) lines, and,
are compared to the case of $\eta$=0 (shown as dotted lines).}
\label{amddbarfs0_dens}
 \end{figure}

\begin{figure}
\vskip -2.in
\includegraphics[width=14cm,height=16cm]{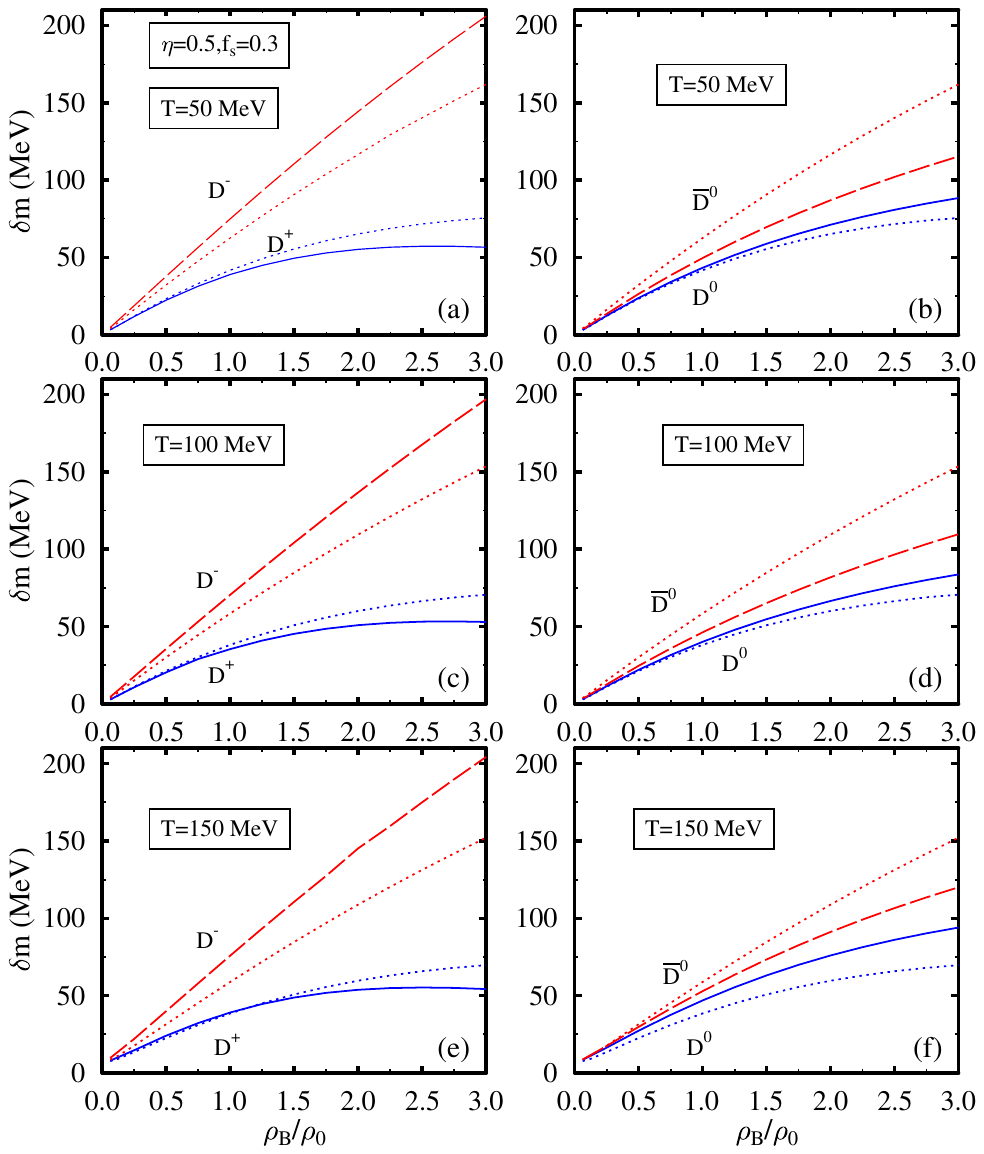} 
\vskip -0.3in
 \caption{
%Same as fig. \ref{amddbarfs0_dens}, with $f_s$=0.3.
Mass shifts of $D^+(D^-)$ (in panels (a), (c) and (e)) 
and $D^{0}(\bar {D^0})$ (in panels (b), (d) and (f)) 
mesons are plotted as functions of the baryon density
in units of nuclear saturation  density
for T=50, 100 and 150 MeV in hyperonic matter (with $f_s$=0.3)
for $\eta=0.5$, as the solid (dashed) lines, and,
are compared to the case of $\eta$=0 (shown as dotted lines).
}
\label{amddbarfs3_dens}
 \end{figure}

\begin{figure}
\vskip -2.in
\includegraphics[width=14cm,height=16cm]{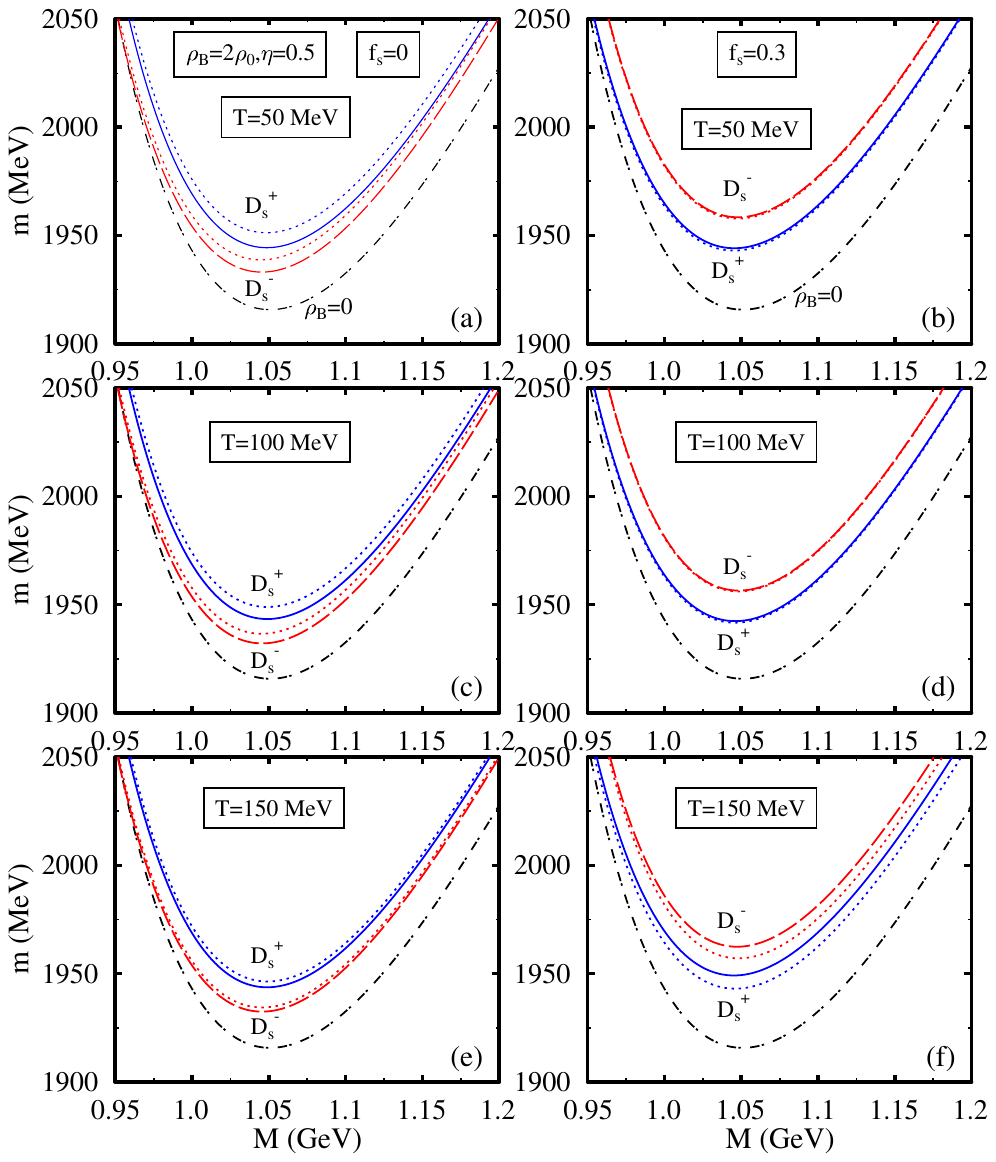} 
\vskip -0.3in
 \caption{
Masses of $D_s^+$ and $D_s^-$ 
mesons are plotted as functions of the Borel mass M 
for T=50, 100 and 150 MeV in panels (a), (c) and (e) 
and in panels (b), (d) and (f), 
in nuclear ($f_s$=0) and hyperonic (with $f_s$=0.3)
matter for baryon density $\rho_B=2\rho_0$ and $\eta=0.5$,
as the solid (dashed) lines,
and, are compared to the case of $\eta$=0 (shown as dotted lines). 
The Borel curve for $\rho_B=0$ is shown as the dot-dashed line.}
        \label{amdspm_2rhb0}
\end{figure}

\begin{figure}
\vskip -2.in
\includegraphics[width=14cm,height=16cm]{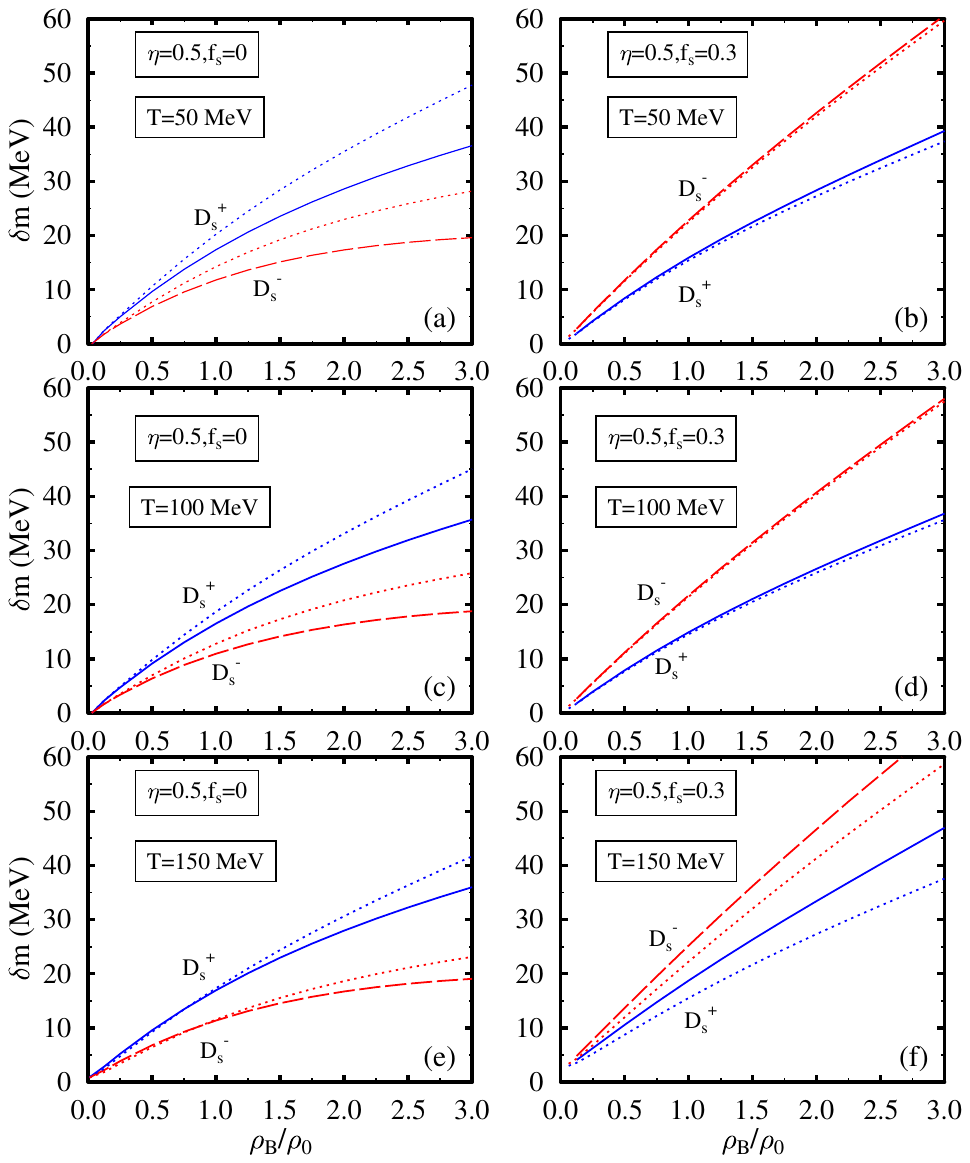} 
\vskip -0.3in
 \caption{
Mass shifts of $D_s^+$ and $D_s^-$ mesons are plotted as functions 
of the baryon density in units of nuclear saturation  density
for $\eta=0.5$ (in panels (a), (c) and (e)) for 
nuclear matter ($f_s$=0) and 
in panels (b), (d) and (f) for strange hadronic matter with
$f_s$=0.3) for T=50, 100 and 150 MeV as the solid
(dashed) lines, and, are
compared to the case of $\eta$=0 (shown as dotted lines). }
\label{amdspm_dens}
 \end{figure}

\begin{figure}
\vskip -2.in
\includegraphics[width=14cm,height=16cm]{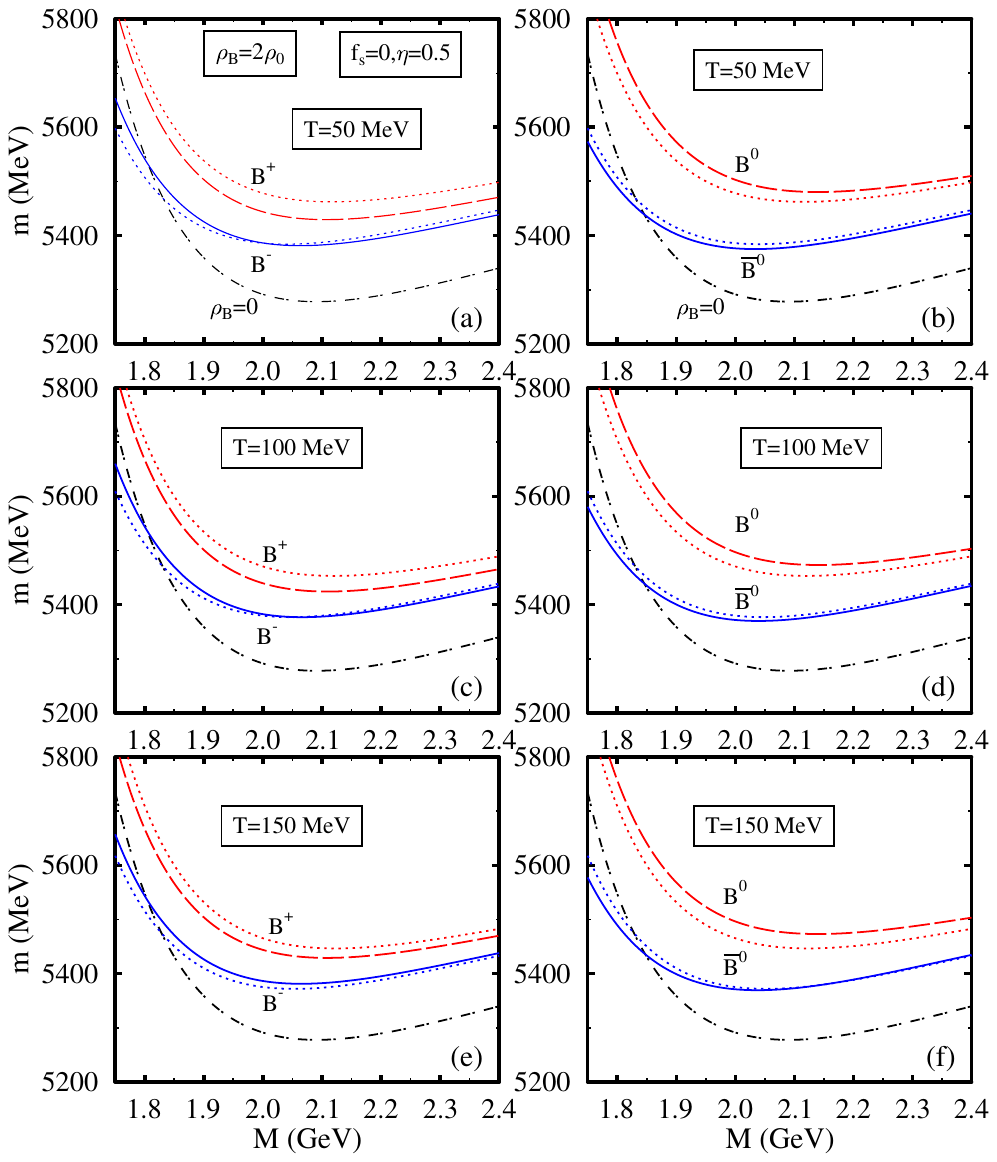} 
\vskip -0.3in
 \caption{
Masses of $B^{-}({B^+})$ (in panels (a), (c) and (e)) 
and $\bar B^0 (B^0)$ (in panels (b), (d) and (f)) 
mesons are plotted as functions of the Borel mass M 
for T=50, 100 and 150 MeV in nuclear matter ($f_s$=0)
for baryon density $\rho_B=2\rho_0$ and $\eta=0.5$,
as the solid (dashed) lines,  
and, are compared to the case of $\eta$=0 (shown as dotted lines). 
The Borel curve for $\rho_B=0$ is shown as the dot-dashed line.}
\label{ambbbarfs0_2rhb0}
 \end{figure}
\begin{figure}
\vskip -2.in
\includegraphics[width=14cm,height=16cm]{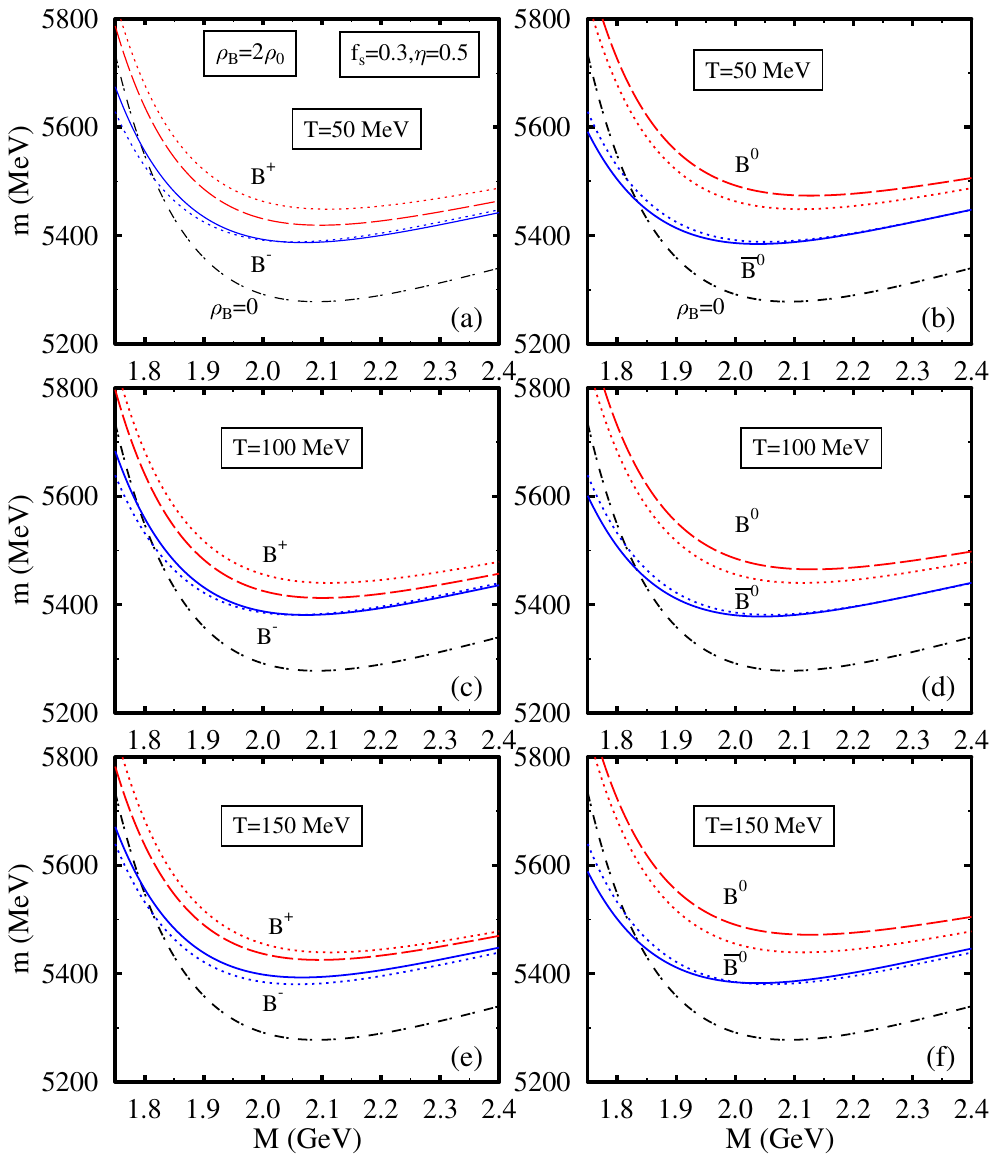} 
\vskip -0.3in
 \caption{
%Same as fig. \ref{ambbbarfs0_2rhb0}, with $f_s$=0.3.
Masses of $B^{-}({B^+})$ (in panels (a), (c) and (e)) 
and $\bar B^0 (B^0)$ (in panels (b), (d) and (f)) 
mesons are plotted as functions of the Borel mass M 
for T=50, 100 and 150 MeV in hyperonic matter (with $f_s$=0.3)
for baryon density $\rho_B=2\rho_0$ and $\eta=0.5$,
as the solid (dashed) lines,  
and, are compared to the case of $\eta$=0 (shown as dotted lines). 
The Borel curve for $\rho_B=0$ is shown as the dot-dashed line.
}
\label{ambbbarfs3_2rhb0}
 \end{figure}

\begin{figure}
\vskip -2.in
\includegraphics[width=14cm,height=16cm]{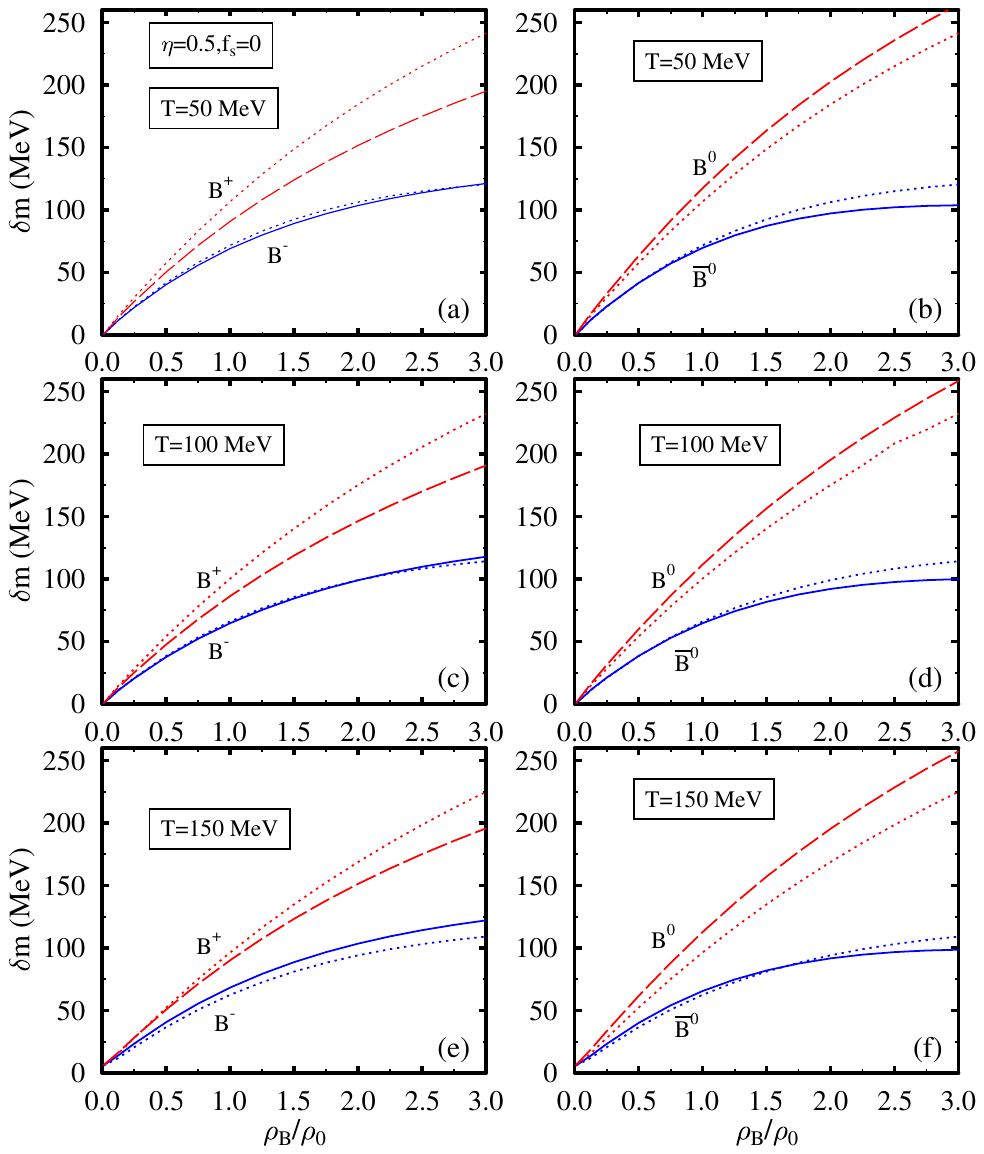} 
\vskip -0.3in
 \caption{
Mass shifts
of $B^{-}({B^+})$ (in panels (a), (c) and (e)) 
and $\bar B^0 (B^0)$ (in panels (b), (d) and (f)) 
mesons are plotted as functions of the baryon density
in units of nuclear saturation  density
for T=50, 100 and 150 MeV in nuclear matter ($f_s$=0)
for $\eta=0.5$,
as the solid (dashed) lines,  
and, are compared to the case of $\eta$=0 (shown as dotted lines). }
\label{ambbbarfs0_dens}
 \end{figure}

\begin{figure}
\vskip -2.in
\includegraphics[width=14cm,height=16cm]{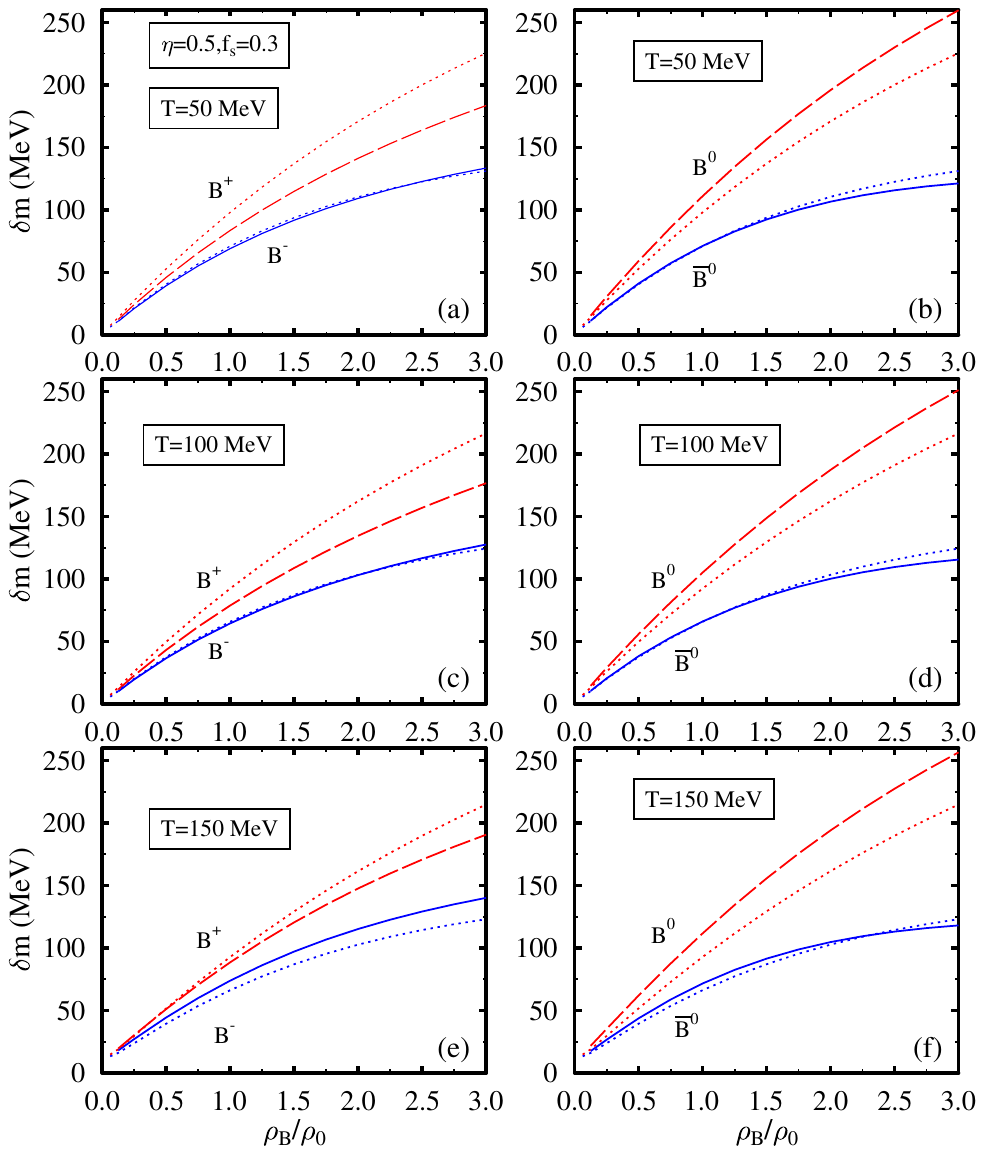} 
\vskip -0.3in
 \caption{
%Same as fig. \ref{ambbbarfs0_dens}, with $f_s$=0.3.
Mass shifts of $B^{-}({B^+})$ (in panels (a), (c) and (e)) 
and $\bar B^0 (B^0)$ (in panels (b), (d) and (f)) 
mesons are plotted as functions of the baryon density
in units of nuclear saturation  density
for T=50, 100 and 150 MeV in hyperonic matter (with $f_s$=0.3)
for $\eta=0.5$, as the solid (dashed) lines,  
and, are compared to the case of $\eta$=0 (shown as dotted lines).
}
\label{ambbbarfs3_dens}
 \end{figure}

\begin{figure}
\vskip -2.in
\includegraphics[width=14cm,height=16cm]{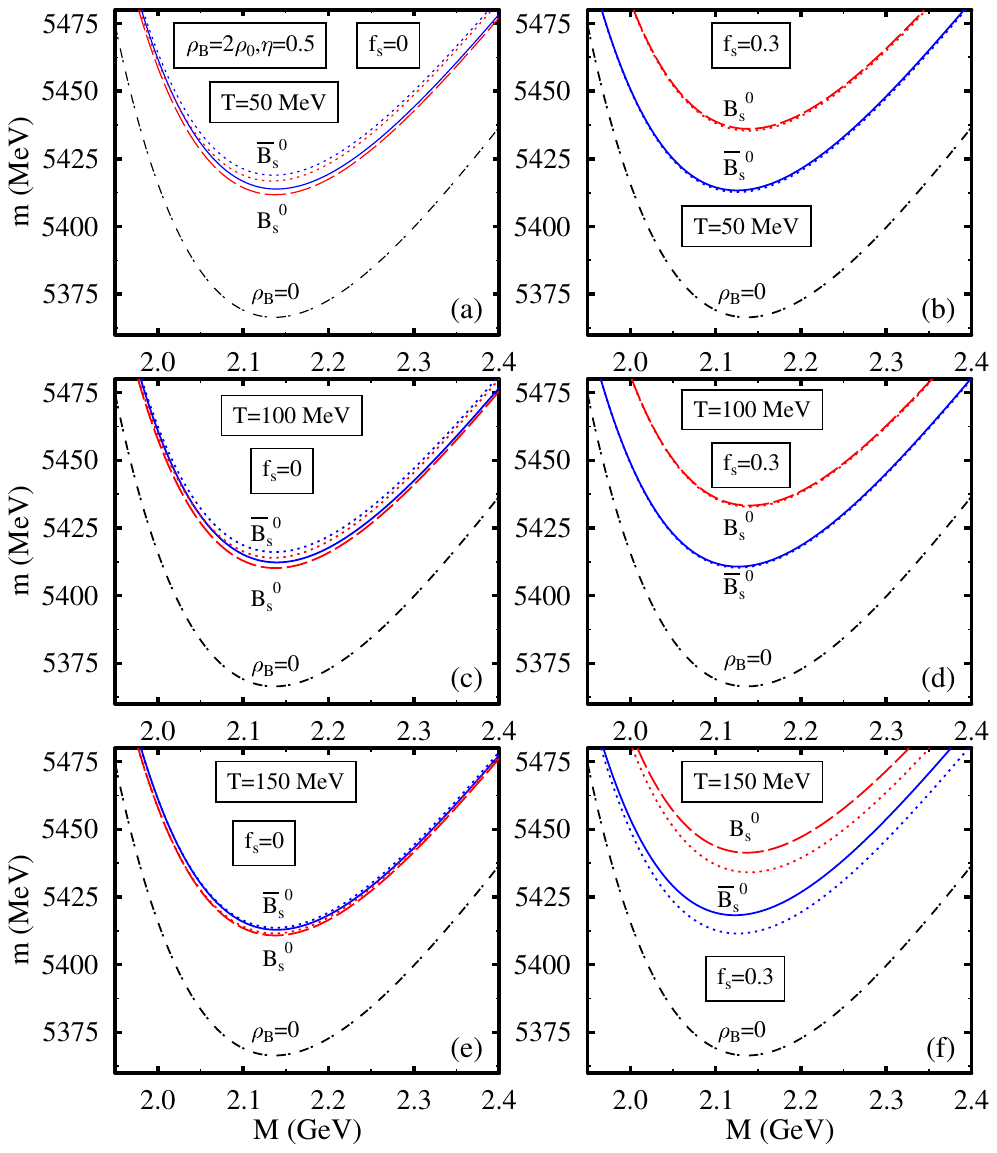} 
\vskip -0.3in
 \caption{
Masses of $\bar {B_s^0}$ and  ${B}_s^0$
mesons are plotted as function of the Borel mass M 
for T=50, 100 and 150 MeV for nuclear matter ($f_s$=0) 
in panels (a), (c) and (e) and 
for hyperonic (with $f_s$=0.3)
in panels (b), (d) and (f), 
for baryon density $\rho_B=2\rho_0$ and $\eta=0.5$,
as the solid (dashed) line,
and, is compared to the case of $\eta$=0 (shown as dotted line). 
The Borel curve for $\rho_B=0$ is shown as the dot-dashed line.}
        \label{ambs0bar_2rhb0}
\end{figure}

\begin{figure}
\vskip -2.in
\includegraphics[width=14cm,height=16cm]{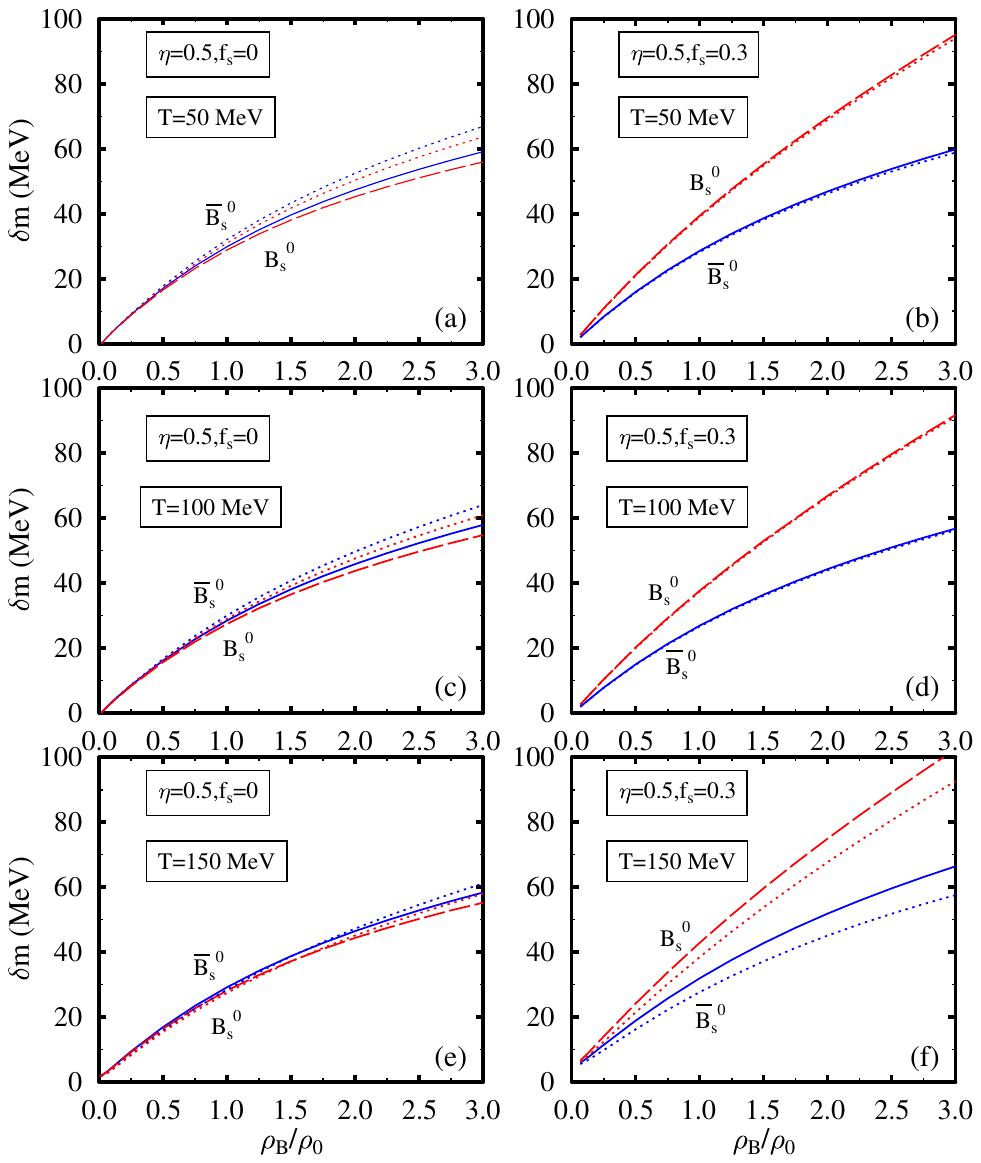} 
\vskip -0.3in
 \caption{
Mass shifts of $\bar {B_s^0}$ and  $B_s^0$
mesons are plotted as function
of the baryon density in units of nuclear saturation  density
for $\eta=0.5$ (in panels (a), (c) and (e)) for 
nuclear matter ($f_s$=0) and 
in panels (b), (d) and (f) for strange hadronic matter with
$f_s$=0.3) for T=50, 100 and 150 MeV 
as the solid (dashed) line,
and, is compared to the case of $\eta$=0 (shown as dotted line). }
\label{ambs0bar_dens}
 \end{figure}

\begin{figure}
\vskip -2.in
\includegraphics[width=14cm,height=16cm]{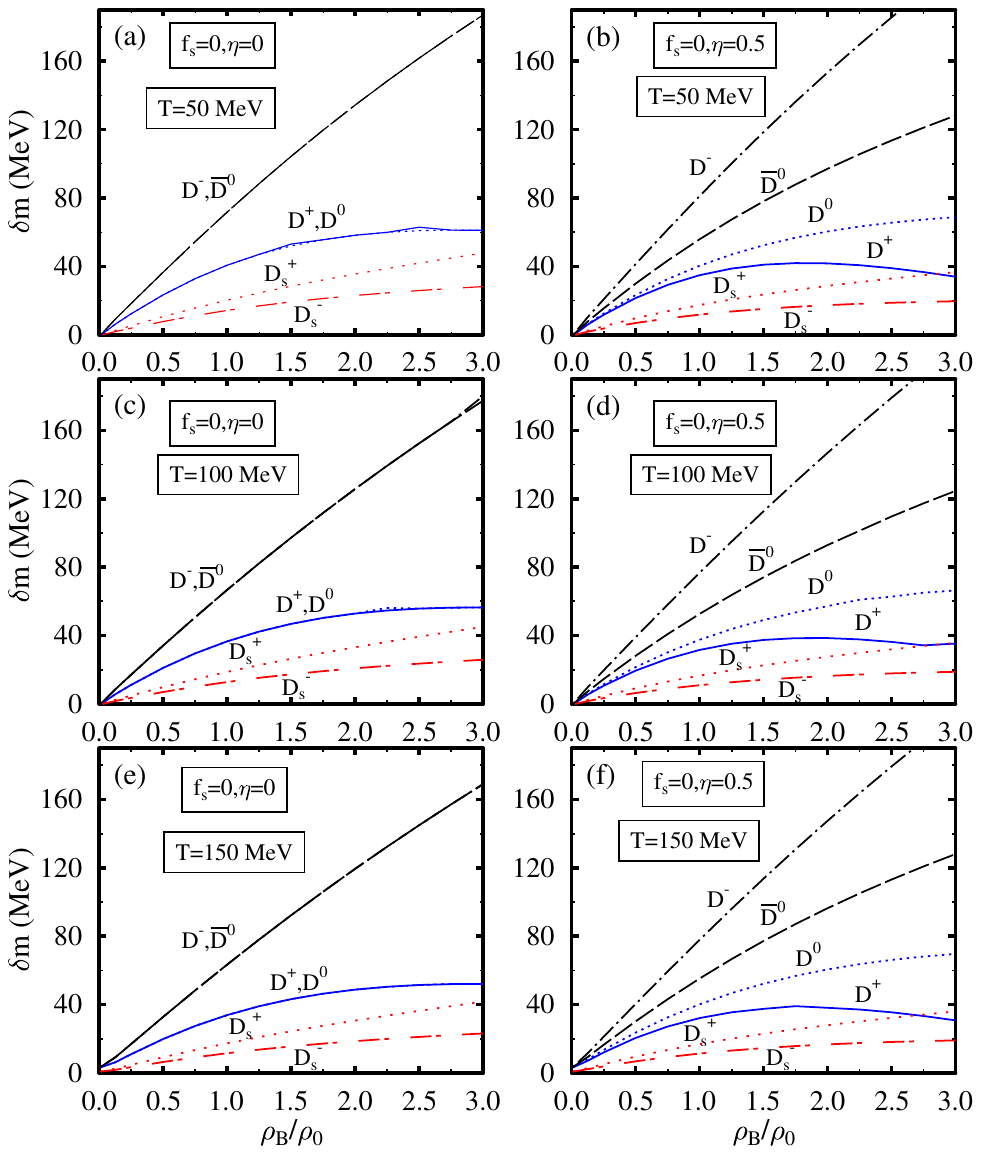} 
\vskip -0.3in
 \caption{Mass shifts of the open charm mesons are plotted 
as a functions of density for the isospin symmetric ($\eta$=0) 
nuclear matter ($f_s$=0) in panels (a), (c) and (e), and,
for the isospin asymmetric (with $\eta$=0.5) 
nuclear matter ($f_s$=0) in panels (b), (d) and (f),
for T=50, 100 and 150 MeV respectively. These mass shifts
are shown as the short-dash-dotted, dashed, closely spaced dotted,
solid, widely spaced dotted, long-dash-dotted   
for the $D^-$, $\bar {D^0}$, $D^0$, $D^+$, $D_s^+$ and $D_s^-$ 
mesons respectively.
}
\label{am_charm_fs0_dens}
 \end{figure}

\begin{figure}
\vskip -2.in
\includegraphics[width=14cm,height=16cm]{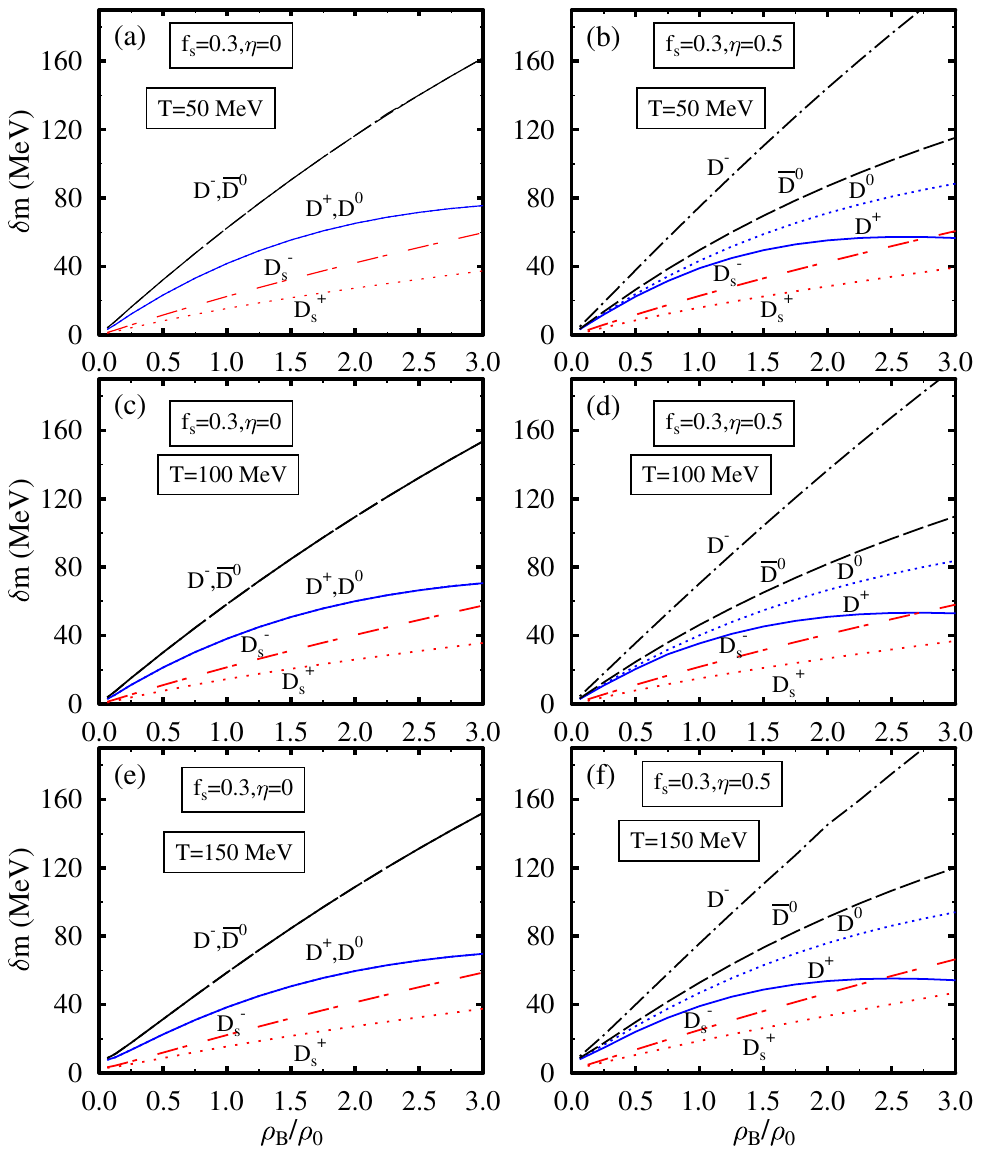} 
\vskip -0.3in
 \caption{
%Same as figure \ref{am_charm_fs0_dens}, with $f_s$=0.3.
Mass shifts of the open charm mesons are plotted 
as a functions of density for the isospin symmetric ($\eta$=0) 
hyperonic matter (with $f_s$=0.3) in panels (a), (c) and (e), and,
for the isospin asymmetric (with $\eta$=0.5) 
hyperonic matter (with $f_s$=0.3) in panels (b), (d) and (f),
for T=50, 100 and 150 MeV respectively. These mass shifts
are shown as the short-dash-dotted, dashed, closely spaced dotted,
solid, widely spaced dotted, long-dash-dotted   
for the $D^-$, $\bar {D^0}$, $D^0$, $D^+$, $D_s^+$ and $D_s^-$ 
mesons respectively.
}
\label{am_charm_fs3_dens}
 \end{figure}

\begin{figure}
\vskip -2.in
\includegraphics[width=14cm,height=16cm]{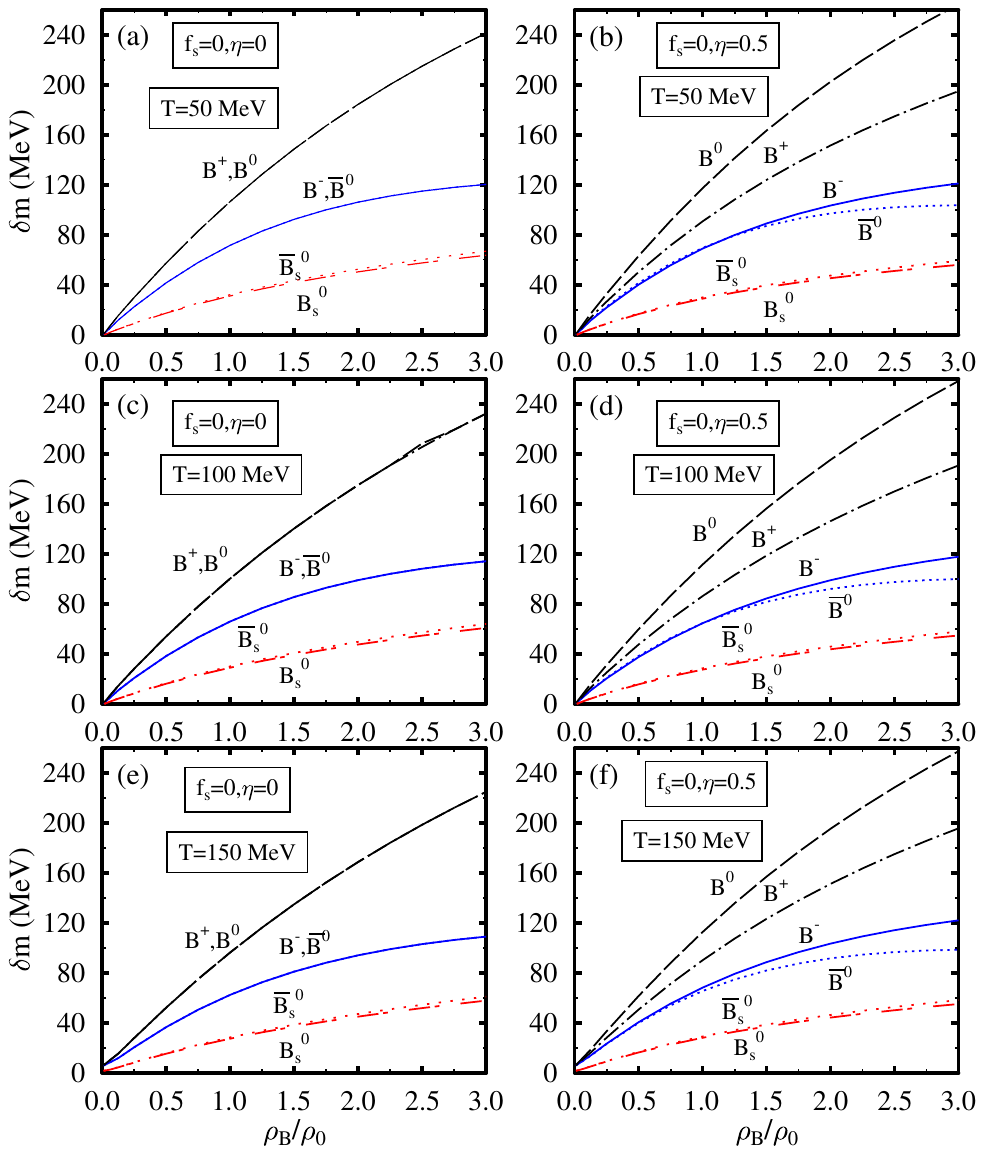} 
\vskip -0.3in
 \caption{Mass shifts of the open bottom mesons are plotted 
as a functions of density for the isospin symmetric ($\eta$=0) 
nuclear matter ($f_s$=0) in panels (a), (c) and (e), and,
for the isospin asymmetric (with $\eta$=0.5) 
nuclear matter ($f_s$=0) in panels (b), (d) and (f),
for T=50, 100 and 150 MeV respectively. These mass shifts
are shown as the short-dash-dotted, dashed, closely spaced dotted,
solid, widely spaced dotted, long-dash-dotted   
for the $B^0$, $B^+$, $\bar {B^0}$, $B^-$, $\bar {B_s^0}$ 
and $B_s^0$ mesons respectively.
}
\label{am_bottom_fs0_dens}
 \end{figure}

\begin{figure}
\vskip -2.in
\includegraphics[width=14cm,height=16cm]{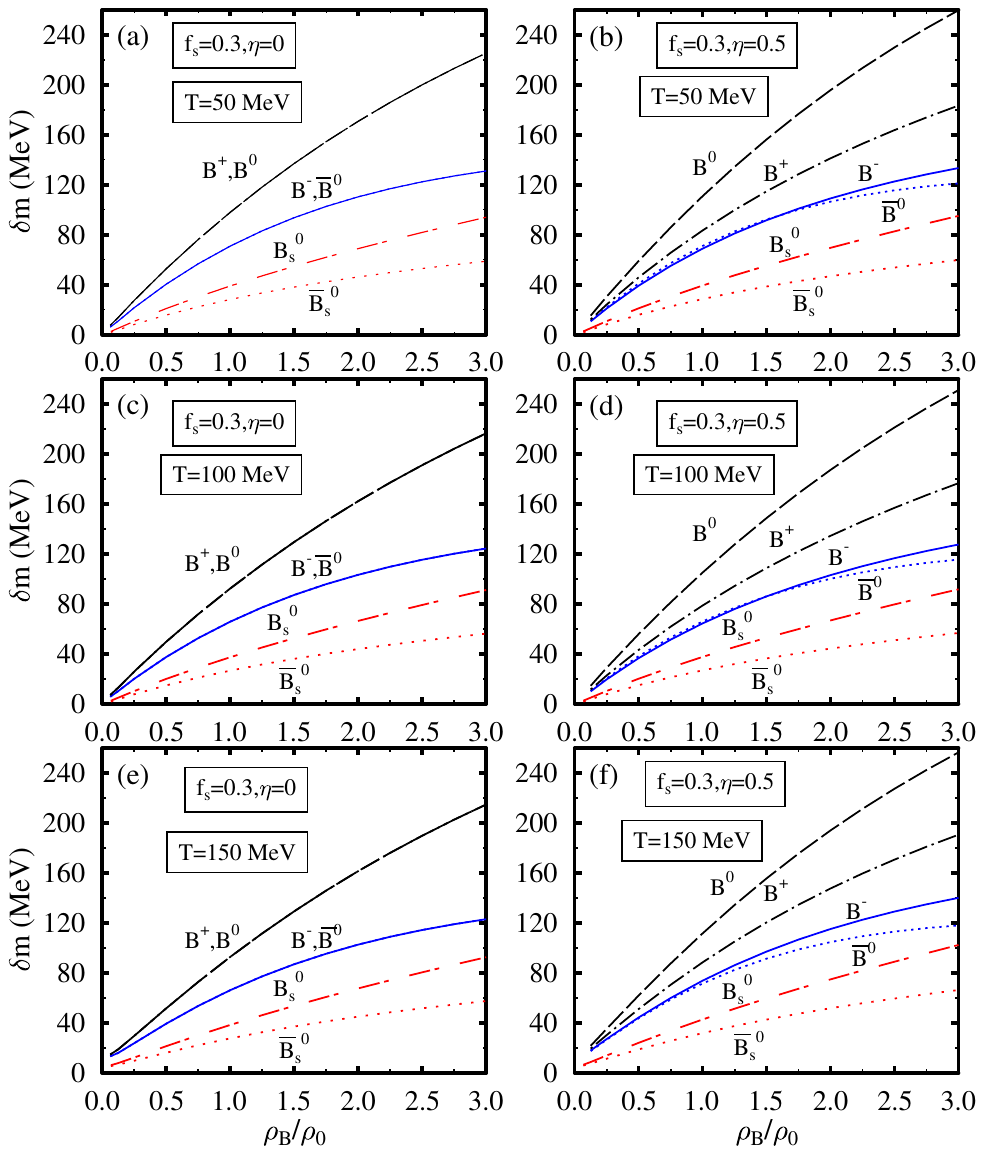} 
\vskip -0.3in
 \caption{
%Same as figure \ref{am_bottom_fs0_dens}, with $f_s$=0.3.
Mass shifts of the open bottom mesons are plotted 
as a functions of density for the isospin symmetric ($\eta$=0) 
hyperonic matter (with $f_s$=0.3) in panels (a), (c) and (e), and,
for the isospin asymmetric (with $\eta$=0.5) 
hyperonic matter (with $f_s$=0.3) in panels (b), (d) and (f),
for T=50, 100 and 150 MeV respectively. These mass shifts
are shown as the short-dash-dotted, dashed, closely spaced dotted,
solid, widely spaced dotted, long-dash-dotted   
for the $B^0$, $B^+$, $\bar {B^0}$, $B^-$, $\bar {B_s^0}$ 
and $B_s^0$ mesons respectively.
}
\label{am_bottom_fs3_dens}
 \end{figure}

The scalar gluon condensate ($\left\langle \frac{\alpha_{s}}{\pi}
G^a_{\mu\nu} {G^a}^{\mu\nu} \right\rangle$
and the expectation value of the twist-2 tensorial 
gluon operator, $\left\langle  \frac{\alpha_{s}}{\pi}
\Big (G^a_{\mu\sigma} {{G^a}_\nu}^{\sigma} u^\mu u^\nu
-\frac{1}{4} G^a_{\mu\sigma} {G^a}^{\mu \sigma}\Big)
\right\rangle$, are obtained from the in-medium value of
the dilaton field, $\chi$ 
\cite{amarvjpsi_qsr},
which is incorporated in the
chiral SU(3) model to simulate the broken scale invariance 
of QCD.
The Lagrangian density for the scalar dilaton field, $\chi$, 
is given as \cite{amarvjpsi_qsr}
\begin{equation}
{\cal L}_\chi  =  \frac {1}{2} (\partial _\mu \chi)(\partial ^\mu \chi)
- k_4 \chi^4 -  \frac{1}{4} \chi^{4} {\rm {ln}} 
\Bigg ( \frac{\chi^{4}} {\chi_{0}^{4}} \Bigg )
+ \frac {d}{3} \chi^{4} {\rm {ln}} \Bigg (\bigg( \frac {\left( \sigma^{2} 
- \delta^{2}\right) \zeta }{\sigma_{0}^{2} \zeta_{0}} \bigg) 
\bigg (\frac {\chi}{\chi_0}\bigg)^3 \Bigg ),
\label{lagchi}
\end{equation}
where, the last two terms in the above Lagrangian density 
are the scale breaking terms.
The energy momentum tensor obtained from the above Langrangian
density is given as
\begin{equation}
T_{\mu \nu}=(\partial _\mu \chi) 
\Bigg (\frac {\partial {{\cal L}_\chi}}
{\partial (\partial ^\nu \chi)}\Bigg )
- g_{\mu \nu} {\cal L}_\chi \equiv
(\partial _\mu \chi)(\partial _\nu \chi) 
- g_{\mu \nu} {\cal L}_\chi.
\label{energymom_chi}
\end{equation}
Accounting for the finite mass of the quarks,  the energy momentum 
of the QCD can be written in terms of a symmetric traceless 
part and a trace part as
%\cite{leemorita2,moritalee_1} 
\begin{equation}
T_{\mu \nu}=-ST({G^a}_{\mu\sigma} {{G^a}_\nu}^{ \sigma})
+ \frac {g_{\mu \nu}}{4} \Big( \frac{\beta_{QCD}}{2g} 
G_{\sigma\kappa}^{a} {G^a}^{\sigma\kappa}
+\sum_i m_i \bar {q_i} q_i \Big),
\label{energymomqcd}
\end{equation}
Writing 
\begin{equation}
\langle \frac {\alpha_s}{\pi}{G^a}_{\mu\sigma} {{G^a}_\nu}^{\sigma} \rangle
=\Big (u_\mu u_\nu - \frac{g_{\mu \nu}}{4} \Big ) G_2,
\label{twist2g2}
\end{equation}
where $u_\mu$ is the 4-velocity of the nuclear medium,
taken as $u_\mu =(1,0,0,0)$, we obtain the energy momentum tensor
in QCD as 
\begin{eqnarray}
T_{\mu \nu}=-\Big (\frac{\pi}{\alpha_s}\Big)\Big (u_\mu u_\nu - 
\frac{g_{\mu \nu}}{4} \Big ) G_2
+ \frac {g_{\mu \nu}}{4} \Big( \frac{\beta_{QCD}}{2g} 
{G^a}_{\sigma \kappa} {G^a}^{\sigma \kappa} 
+\sum_i m_i \bar {q_i} q_i \Big).
\label{energymomqcdg2}
\end{eqnarray}
We calculate the expectation values of the scalar and twist-2 
gluon operators by equating the energy momentum tensor 
for the dilaton field with that of QCD, multiplying both sides by 
$g^{\mu \nu}$ and $(u^\mu u^\nu -\frac {g^{\mu \nu}}{4})$ 
($u^\mu = (1, 0,0,0)$ is the 4-velocity of the hadronic medium at rest)
respectively, and summing over the indices ($\mu$ and $\nu$)
\cite{amarvjpsi_qsr}. We then obtain \cite{amarvjpsi_qsr}
\begin{equation}
{T_{\mu}}^{\mu} = \sum_i m_i \bar {q_i} q_i+ \langle \frac{\beta_{QCD}}{2g} 
G_{\mu\nu}^{a} G^{\mu\nu a} \rangle = -(1 - d)\chi^{4} 
\label{tensor2m}
\end{equation}
and, 
\begin{eqnarray}
&& \langle \frac {\alpha_s}{\pi}{G^a}_{\mu\sigma} {{G^a}_\nu}^{\sigma} \rangle
(u^\mu u^\nu -\frac {g^{\mu \nu}}{4})\equiv \frac{3}{4} G_2
=\frac {3\alpha_s}{4\pi}
\Bigg [-(1-d+4 k_4)(\chi^4-{\chi_0}^4)\nonumber \\ 
&-&\chi ^4 {\rm {ln}}
\Big (\frac{\chi^4}{{\chi_0}^4}\Big )
 +  \frac {4}{3} d\chi^{4} {\rm {ln}} \Bigg (\bigg( \frac {\left( \sigma^{2} 
- \delta^{2}\right) \zeta }{\sigma_{0}^{2} \zeta_{0}} \bigg) 
\bigg (\frac {\chi}{\chi_0}\bigg)^3 \Bigg ) \Bigg ].
\label{twist2approx}
\end{eqnarray}
In equation (\ref{tensor2m}), the first term of the trace 
of the energy-momentum tensor within the chiral 
SU(3) model is the negative of the explicit chiral symmetry breaking
term, ${\cal L}_{SB}$ given by equation (\ref{lag_SB_MFT}).
It might be noted here that in deriving the relation 
of the twist-2 gluon operator with the dilaton field
(given by eqn (\ref{twist2approx})),we have assumed that 
the dilaton field, $\chi$ is non-relativistic, and hence
have assumed that $\langle (\partial_i \chi) 
(\partial _i \chi)\rangle_{medium}
\simeq \langle (\partial_i \chi) (\partial _i \chi)\rangle_{vac}$.
Using the QCD $\beta$ function at one loop level, for 
$N_{c}$ colors and $N_{f}$ flavors 
\begin{equation}
\beta_{\rm {QCD}} \left( g \right) = -\frac{11 N_{c} g^{3}}{48 \pi^{2}} 
\left( 1 - \frac{2 N_{f}}{11 N_{c}} \right),
\label{beta}
\end{equation}
for $N_c$=3 and $N_f$=3, the scalar gluon condensate 
is obtained as
\begin{equation}
\left\langle  \frac{\alpha_{s}}{\pi} {G^a}_{\mu\nu} {G^a}^{\mu\nu} 
\right\rangle =  \frac{8}{9} \Bigg [(1 - d) \chi^{4}
+\left( \frac {\chi}{\chi_{0}}\right)^{2} 
\left( m_{\pi}^{2} f_{\pi} \sigma
+ \big( \sqrt {2} m_{k}^{2}f_{k} - \frac {1}{\sqrt {2}} 
m_{\pi}^{2} f_{\pi} \big) \zeta \right) \Bigg ]. 
\label{chiglu}
\end{equation}
The parameter $d$ in equation (\ref{tensor2m}) originates from the 
second logarithmic term of equation (\ref{lagchi}). 
The parameter $d$, along with the other parameters
corresponding to the meson-meson interaction Lagrangian density,
${\cal L}_0$ of the general Lagrangian 
given by (\ref{genlag}), are fitted so as to ensure 
extrema in the vacuum for the $\sigma$, $\zeta$ and $\chi$ field 
equations, to  reproduce the vacuum masses of the $\eta$ and $\eta '$ 
mesons, the mass of the $\sigma$ meson around 500 MeV, and,
pressure, p($\rho_0$)=0,
with $\rho_0$ as the nuclear matter saturation density 
\cite{paper3,amarindamprc}.
The scalar fields (the non-strange isoscalar field, $\sigma$, 
the strange isoscalar field, $\zeta$ 
and the isovector field $\delta$), the dilaton field, $\chi$ 
and the vector fields (non-strange isoscalar
$\omega$, strange isoscalar $\phi$ and non-strange isovector $\rho$) 
are obtained by minimizing 
the thermodynamic potential $\Omega$ for
given values of temperature, baryonic density, $\rho_B$, 
the isospin asymmetry parameter, 
$\eta = -\frac{\Sigma_i I_{3i} \rho_{i}}{\rho_{B}}$, the strangeness 
fraction, $f_s = \frac{\Sigma_i \vert s_{i} \vert \rho_{i}}{\rho_{B}}$,
where, $I_{3i}$ and $\vert s_{i} \vert$ denote the third component 
of isospin quantum number and the number of strange quarks 
in the $i^{th}$ baryon, respectively.

The effects of the scalar and twist-2
gluon condensates, as obtained from the medium modification
of the dilaton field, $\chi$, within the chiral SU(3) model,
on the masses of the vector and pseudoscalar charmonium 
ground states ($J/\psi$ and $\eta_c$)
were studied in Ref. \cite{amarvjpsi_qsr}
and on the masses of the heavy quarkonium 
(charmonium and bottomonium)
states in Ref. \cite{QCDSR_Heavy_Quarkonia}, 
using QCD sum rule approach. 
In the present work, the effects of the medium modifications
of the QCD operators, upto dimension 5, on the masses of the
open charm and open bottom mesons are investigated
within the QCD sum rule approach. 
The non-strange ($\langle \bar u u\rangle$, $\langle \bar d d\rangle$) 
and strange  ($\langle \bar s s\rangle$) quark condensates
are obtained from the scalar fields using equation (\ref{qqbar}) 
and the scalar and twist-2 gluon condensates
are calculated from the dilaton field, $\chi$, 
using equations (\ref{tensor2m} and (\ref{twist2approx})
within the chiral SU(3) model. 
The dimension 5 mixed quark-gluon condensate 
$\langle \bar {q_i} g_s \sigma . G {q_i} \rangle$
in the operator product expansion is obtained from the in-medium
quark condensate $\langle \bar {q_i} {q_i} \rangle$,
with $i=1,2,3\equiv u,d,s$,
as computed within the chiral SU(3) model, 
by using the relation 
\begin{equation}
\langle \bar {q_i} g_s \sigma . G {q_i} \rangle=
2\lambda_{q}^2 \langle \bar {q_i} {q_i} \rangle.
\label{mixed_quark_gluon_condensate_lambda_q}
\end{equation}
The values of the parameter $\lambda_q$ estimated using
different methods are, however, observed to be widely different
\cite{Cohen_Prog_Part_Nucl_Phys,Belyaev,lambda_q2_lat_1,lambda_q2_lat_2,Kremer,Mikhailov,Shuryak}.
Using the leading order approximations for the parton
distribution functions \cite{Glueck_Reya_Vogt}, 
the values of the higher order operators (or combinations
of these operators) used in the present investigation are as follows 
\cite{Jin_Cohen_Furnstahl_Griegel_PRC47_2882_1983}.
The estimates for $q_i=u,d$ are 
\begin{equation}
\langle q_i^\dagger iD_0 q_i\rangle=0.18 \; {\rm GeV} \rho_B,
\end{equation}
\begin{eqnarray}
-\langle {\bar q_i} D_0^2 q_i\rangle
+\frac{1}{8}\langle \bar {q_i} g_s \sigma . G {q_i} \rangle
%= -\langle {\bar q_i} D_0^2 q_i\rangle_{med}
%+\frac{1}{8}\langle \bar {q_i} g_s \sigma . G {q_i} \rangle_{med}
= 0.3 \;{\rm GeV}^2 \; \rho_B,
\label{F1_comb}
\end{eqnarray}
and
\begin{eqnarray}
-\langle q_i^\dagger D_0^2 q_i\rangle
+\frac{1}{12}\langle {q_i}^\dagger g_s \sigma . G {q_i} \rangle
%= -\langle q_i^\dagger D_0^2 q\rangle_{med}
%+\frac{1}{12}\langle {q_i}^\dagger g_s \sigma . G {q_i} \rangle_{med}
= 0.031\; {\rm GeV}^2 \; \rho_B,
\label{F2_comb}
\end{eqnarray}
for the case of symmetric nuclear matter. 
For the strange quark ($q_i=s$), the estimate for the above operator 
is taken to be $\langle s^\dagger iD_0 s\rangle = 
0.06 \; {\rm GeV} \rho_B$
\cite{Cohen_Prog_Part_Nucl_Phys}. 
For the consideration of the strange-heavy flavor mesons,
the estimates of the combinations of the operators
given by equations (\ref{F1_comb}) and (\ref{F2_comb})
(for $q_i$=s) are given as \cite{open_heavy_flavour_qsr_2}
\begin{eqnarray}
-\langle {\bar s} D_0^2 s\rangle
+\frac{1}{8}\langle \bar {s} g_s \sigma . G {s} \rangle
= y
\Big(-\langle q_i^\dagger D_0^2 q_i\rangle
+\frac{1}{12}\langle {q_i}^\dagger g_s \sigma . G {q_i} \rangle
\Big)
\label{F1_comb_s}
\end{eqnarray}
and
\begin{eqnarray}
-\langle s^\dagger D_0^2 s\rangle
+\frac{1}{12}\langle {s}^\dagger g_s \sigma . G {s} \rangle
=y \Big(-\langle q_i^\dagger D_0^2 q_i\rangle
+\frac{1}{12}\langle {q_i}^\dagger g_s \sigma . G {q_i} \rangle
\Big),
\label{F2_comb_s}
\end{eqnarray}
where, $q_i=u\;{\rm {or}}\; d$ in the RHS of the above equations, 
which have equal values for the operators for symmetric nuclear matter.
In equations (\ref{F1_comb_s}) and  (\ref{F2_comb_s}),
$y$ is the strangeness content of the nucleon, given as  
\begin{equation}
y={\langle \bar s s \rangle_{med}}/{\langle \bar q q \rangle_{med}},
\label{y}
\end{equation}
with $\langle \bar s s\rangle_{med}=\langle \bar s s\rangle
-\langle \bar s s\rangle_{vac}$, and,
${\langle \bar q q \rangle}_{med}=({\langle \bar u u \rangle}_{med}+
{\langle \bar d d \rangle}_{med})/2$ is the average of the 
quark-antiquark condensates for $u$ and $d$ quarks after subtracting the
vacuum contributions.
For the isospin asymmetric nuclear (hyperonic) matter
as considered in the present investigation,
we generalize the above relations as follows.
For the non-strange quarks, $q_i=u,d$, for $i=1,2$,
\begin{equation}
-\langle {\bar q_i} D_0^2 q\rangle
+\frac{1}{8}\langle \bar {q_i} g_s \sigma . G {q_i} \rangle
= 0.3 \;{\rm GeV}^2 \; 
\bigg(\frac {\langle \bar {q_i} q_i\rangle_{med}}{\langle \bar {q} q\rangle_{med}}\bigg)\;
\rho_B, 
\label{F1_comb_asym}
\end{equation}
\begin{equation}
-\langle q_i^\dagger D_0^2 q_i\rangle
+\frac{1}{12}\langle {q_i}^\dagger g_s \sigma . G {q_i} \rangle
= 0.031 \;{\rm GeV}^2 \; 
\bigg(\frac {\langle \bar {q_i} q_i\rangle_{med}}{\langle \bar {q} q\rangle_{med}}\bigg)\;
\rho_B,
\label{F2_comb_asym}
\end{equation}
For the QCD operators involving the strange quark, we take 
\begin{equation}
-\langle {\bar s} D_0^2 s\rangle
+\frac{1}{8}\langle \bar {s} g_s \sigma . G {s} \rangle
= 0.3 \;{\rm GeV}^2 \; \rho_B\; y,
\label{F1_comb_s_asym}
\end{equation}
\begin{equation}
-\langle s^\dagger D_0^2 s\rangle
+\frac{1}{12}\langle {s}^\dagger g_s \sigma . G {s} \rangle
= 0.031 \;{\rm GeV}^2 \; \rho_B\; y,
\label{F2_comb_s_asym}
\end{equation}
where $y$ is the strangeness content of the nucleon defined
by equation (\ref{y}).
The estimate for the operator 
$\langle q_i^\dagger g_s \sigma . G q_i\rangle
=0.33 \;{\rm {GeV}}^2 \rho_B$, for $q_i=u\;\; {\rm {or}}\;\; d$ 
for the symmetric nuclear matter \cite{open_heavy_flavour_qsr_2}
is generalized to the isospin asymmetric case in the 
present investigation as
\begin{equation}
\langle q_i^\dagger g_s \sigma . G q_i\rangle
=0.33 \;{\rm {GeV}}^2 \;
\bigg(\frac {\langle \bar {q_i} q_i\rangle_{med}}{\langle \bar {q} q\rangle_{med}}\bigg)\;
\rho_B, 
\label{qidgsgGqi}
\end{equation}
for $q_i=u\;\; {\rm {and}}\;\; d$, and, 
for the operator with strange quark given as,
\begin{equation}
\langle {s}^\dagger g_s \sigma . G {s} \rangle
= 0.33 \;{\rm GeV}^2 \; \rho_B\; y.
\label{sdgsgGs}
\end{equation}
Using the estimates for the combinations of the operators
given by equations (\ref{F2_comb_asym}) and (\ref{F2_comb_s_asym}),
one obtains the estimates for 
$\langle {q_i}^\dagger D_0^2 {q_i}\rangle$, 
for $(q_i,i=1,2,3) \equiv (u,d,s)$.
Using the values for the QCD operators for the even and odd parts
of the spectral function as obtained above, the masses
of the open charm and open bottom mesons are computed within
the QCD sum rule approach.

\section{Results and discussions}

In this section, we discuss the results obtained
for the in-medium masses of the open charm 
($D$, $\bar D$, $D_s$ and $\bar {D_s}$) 
and open bottom ($B$, $\bar B$, $B_s$ and $\bar {B_s}$) 
mesons, and, their antiparticles in asymmetric 
strange hadronic matter at finite temperatures.
These are computed using the QCD sum rule approach 
retaining QCD operators upto dimension 5, with the 
medium dependence, e.g., the temperature, isospin asymmetry,
strangeness and density dependence of these operators
calculated using a chiral SU(3) model.
In the present investigation, we choose the quark mass parameters 
to be $m_u=m_d=6\; {\rm  MeV}$, $m_s$=185 MeV, 
$m_c$=1.5 GeV and $m_b$=4.7 GeV. The light quark condensates
($\langle \bar q_i q_i\rangle$, $(i=1,2,3)\equiv (u,d,s)$)
are obtained from the values of the scalar fields, $\sigma$,
$\zeta$ and $\delta$ in the chiral SU(3) model 
by using equation (\ref{qqbar}).
For the chosen values of the masses of the quarks, 
the vacuum values of the light quark condensates are obtained
as $\langle \bar q q \rangle_{vac}=\langle \bar u u \rangle_{vac}
=\langle \bar d d \rangle_{vac}=
(-241.1\; {\rm MeV})^3 \; {\rm and}\; \langle \bar s s \rangle_{vac}=
(-288.1 \; {\rm MeV })^3$. The non-strange quark condensate,
$\langle \bar q q \rangle_{vac}$ may be compared with the 
value of $(-225\; {\rm MeV})^3$ 
\cite{Jin_Cohen_Furnstahl_Griegel_PRC47_2882_1983}
and the average estimate of $(-272\; {\rm MeV})^3$ 
of Flavor Lattice Averaging group (FLAG) \cite{FLAG}. 
As may be observed, for the strange quark condensate, 
the value of $(-288.1 \; {\rm MeV })^3$ in the present investigation 
has close agreement with the estimates of 
$\langle \bar s s \rangle_{vac}=(-290 \; {\rm MeV })^3$ 
\cite{lat_sbars_1} and 
$\langle \bar s s \rangle_{vac}=(-296 \; {\rm MeV })^3$ 
\cite{lat_sbars_2}  from lattice calculations.
The value of $\langle \bar s s \rangle_{vac}$
of $(-288.1 \; {\rm MeV })^3$ in the present
work as well as from the lattice calculations
\cite{lat_sbars_1,lat_sbars_2} are much higher
than the value $\langle \bar s s \rangle_{vac}=
(-208 \; {\rm MeV })^3$ obtained using the parametrization
$\langle \bar s s \rangle=0.8 \langle \bar q q \rangle$
used in Ref. \cite{Jin_Cohen_Furnstahl_Griegel_PRC47_2882_1983}.
In the present work, the light quark condensates 
in vacuum are observed to be modified to the values 
$\langle \bar u u \rangle=\langle \bar d d \rangle=
(-206.1 \; {\rm MeV} )^3 \; {\rm and}\; \langle \bar s s \rangle=
(-277.3\; {\rm MeV} )^3$ in symmetric nuclear matter 
at nuclear matter saturation density ($\rho_0$).
The strangeness content nucleon, $y$, as defined in equation
(\ref{y}) has a value of 0.49 for $\rho_B=\rho_0$
in the chiral model as used in the present investigation.
The strangeness content, $y$ is related to the 
the nucleon $\sigma$-term as 
\cite{Cohen_Prog_Part_Nucl_Phys,Gasser_etal_PLB253_1991_252}
\begin{equation}
y=\frac{\sigma_N^0}{\sigma_N}-1,
\label{y_nucleon_sigma_term}
\end{equation} 
with the nucleon $\sigma$ term defined as
\begin{equation}
\sigma_N=(m_u+m_d)\frac{\big(\langle \bar q q \rangle_{med}-
\langle \bar q q \rangle_{vac}\big)}{\rho_B},
\label{nucleon_sigma_term}
\end{equation} 
In equation (\ref{y_nucleon_sigma_term}), $\sigma_N^0$ is
the nucleon $\sigma$ term, in the limit of strangeness content
of the nucleon to be zero. Using the estimates for $\sigma_N$
and $\sigma_N^0$ as 45 and 35 MeV respectively
\cite{Cohen_Prog_Part_Nucl_Phys,Gasser_etal_PLB253_1991_252}, 
the value of $y$ is obtained
to be 0.286 at the nuclear matter saturation density.
The value for $y$ estimated using lattice calculations
in Ref. \cite{Dong_etal_PRD54_5496_1996} is 0.36. 
The value of $y$ is thus not well estimated,
and is varied within the range $0-0.5$ \cite{Navarra}.
We might note here that the value of the nucleon $\sigma$ term
(using the equation (\ref{nucleon_sigma_term})) for the obtained
value of $y=0.49$ in the present investigation, is calculated 
to be $\sigma_N=54.8 \;{\rm MeV}$. This value of the present 
work may be compared to the often used value of 45 MeV 
\cite{Cohen_Prog_Part_Nucl_Phys}. 

In the present work, the value for the scalar gluon condensate, 
$\left\langle \frac{\alpha_{s}}{\pi} 
G^a_{\mu\nu} {G^a}^{\mu\nu} \right\rangle$, 
obtained in vacuum is $(373.1\; {\rm  MeV})^4=0.019\; {\rm GeV}^4$,
which is modified to $(370.6\; {\rm  MeV})^4=0.0189 \; {\rm GeV}^4$ 
at $\rho_B=\rho_0$ in symmetric nuclear matter.
The value of $0.019 \;{\rm GeV}^4$ in vacuum may be compared 
with the lattice results of $0.028 \;{\rm GeV}^4$
\cite{gg_lat_1} and $0.077 \;{\rm GeV}^4$ \cite{gg_lat_2}.
The estimate of the scalar gluon condensate is thus observed
to be larger in the lattice calculations as well as present work,
as compared to the value of $0.012 \;{\rm GeV}^4$
obtained in QCD sum rule calculations in Ref. \cite{SVZ}.
As has already been mentioned, the parameter 
$\lambda_q^2$ relating the mixed quark-gluon condensate
to the quark condensate (given by equation
(\ref{mixed_quark_gluon_condensate_lambda_q}))
is not well estimated \cite{Cohen_Prog_Part_Nucl_Phys}. 
The estimates for $\lambda_q^2$ in vacuum
are 0.4 GeV$^2$ using standard QCD sum rule approach 
\cite{Belyaev}, 1.25 GeV$^2$ \cite{lambda_q2_lat_1}
and $0.5-0.55$ GeV$^2$ \cite{lambda_q2_lat_2,Kremer} 
estimated using lattice QCD, and, $\lambda_q^2 \sim$ 1.2 GeV$^2$,
in a QCD sum rule analysis of the pion wave function using
non-local condensates \cite{Mikhailov} 
as well as within an instanton liquid model \cite{Shuryak}.
In the present work, we take the value of this parameter
to be $\lambda_q^2$=1.2\; GeV$^2$.
For the charm sector,
the vacuum masses for the $D (D^0,D^+)$ and $\bar D (\bar {D^0},D^-)$ 
mesons are obtained 
to be 1869 MeV for ${\tilde s}_0$=7.5 GeV$^2$, and, the mass of $D_s^\pm$ 
is obtained to be around 1916 MeV for ${\tilde s}_0$=15 GeV$^2$.
For the bottom sector,
the vacuum masses for the $B (B^+,B^0)$ and $\bar B (B^-,\bar {B^0}$ 
mesons are obtained 
to be 5277.8 MeV for ${\tilde s}_0$=37.8 GeV$^2$, and, the mass of 
$B_s^0 ({\bar B}_s^0)$ 
is obtained to be around 5366.455 MeV for ${\tilde s}_0=42.3\; {\rm {GeV}}^2$.
There is a splitting between the masses of the open heavy flavor
particles ($D^0$, $D^+$, $D_s^+$, $B^-$, $\bar {B^0}$ and ${\bar {B_s}}^0$) 
and their antiparticles ($\bar {D^0}$, ${D^-}$, $D_s^-$,
$B^+$, $B^0$ and $B_s^0$), which arises from the contributions
from the number density of the $i$-th quark 
($\langle {q_i}^\dagger {q_i}\rangle$),
along with the contributions from the dimension 5 operators,  
$\langle \bar {q_i}D_0^2 {q_i}\rangle$ and
$\langle {q_i}^\dagger g_s \sigma \cdot G q_i\rangle$  of the
odd part of the spectral function. 

In figures \ref{amddbarfs0_2rhb0}
and \ref{amddbarfs3_2rhb0}, we plot the
masses of the charged and neutral open charm mesons 
as functions of the Borel mass, $M$, for the cases of 
nuclear matter ($f_s$=0) and strange hadronic matter
(with $f_s$=0.3) for $\rho_B=2\rho_0$ and T=50, 100 and 150 MeV. 
These are shown for the isospin asymmetry parameter, $\eta$=0.5, and
are compared to the case of isospin symmetric case ($\eta$=0).
The vacuum case is shown as the dot-dashed line.
At finite densities, shown for $\rho_B=2\rho_0$
in these figures, the mass of the antiparticle $D^-(\bar{D^0}$) 
is observed to be higher than $D^+(D^0)$ both in the hot 
nuclear matter and strange hadronic matter. 
As can be observed from these figures,
the effect of isospin asymmetry is to reduce (increase) the mass 
of $D^+(D^0)$ within the iso-doublet ($D^0$, $D^+$), and,
of ${\bar D}^0 (D^-)$ within the iso-doublet ($D^-$,${\bar D}^0$)
for both nuclear matter as well as in strange hadronic matter.
The effect from isospin asymmetry is observed to be much smaller
for the neutral particle $D^0$ as compared to the effects on its 
antiparticle $\bar {D^0}$. 

The shifts in the masses of the charged ($D^\pm$) and neutral 
($D^0, {\bar D}^0$) from the vacuum values 
are plotted as functions of the baryon density in units of
nuclear matter saturation density, in figures 
\ref{amddbarfs0_dens} and \ref{amddbarfs3_dens}  for 
nuclear matter ($f_s$=0) and strange hadronic matter 
(with $f_s$=0.3) respectively.
The effect from isospin asymmetry is observed to be appreciable
for higher values of the baryon density, especially for
$\bar D^0$ meson.
In isospin symmetric ($\eta$=0) nuclear (hyperonic) matter,
the masses of open charm mesons within the $D (D^0,D^+)$ 
and $\bar D (D^-,\bar {D^0})$ doublets are degenerate. 
At $\rho_B=\rho_0$, the difference in the mass of the 
particle $D^0(D^+)$ 
($\sim$ 1912 MeV) and the mass of its antiparticle $\bar {D^0}(D^-)$ 
($\sim$ 1944 MeV) is around  
32 MeV, which may be compared to the mass difference of
around 30 MeV in Ref.  \cite{open_heavy_flavour_qsr_2}
for the isospin symmetric ($\eta$=0) nuclear matter. 
In the presence of strangeness in the medium (with $f_s$=0.3), 
as plotted in figure \ref{amddbarfs3_dens},
for isospin symmetric matter ($\eta$=0), 
the mass difference of $D^0(D^+)$ ($\sim$ 1914.4 MeV)
and $\bar {D^0}(D^-)$ ($\sim$ 1935.9 MeV) is observed 
to be modified to 21.5 MeV.
In the presence of isospin asymmetry, the mass splitting 
between the charged (neutral) particle and antiparticle, 
$D^+ (D^0)$ and $D^-(\bar {D^0})$ is observed to
increase (drop) as the density is raised.
However, the effect from the ispspin asymmetry is lessened
in the presence of strangeness in the medium, as can be
seen from figures \ref{amddbarfs0_dens} and \ref{amddbarfs3_dens}.

In figure \ref{amdspm_2rhb0}, the masses of the strange-charm mesons,
$D_s^\pm$ are plotted as functions of the Borel mass, $M$ for the nuclear
and strange hadronic matter for $\rho_B=2\rho_0$. The Borel curves
for these masses are shown for the vacuum (shown as a dot-dashed line).
In nuclear matter, unlike the observation of the mass of the particle, 
$D^+(D^0)$ to be smaller than the mass of the antiparticle, 
$D^-({\bar D}^0)$ (as seen in figures \ref{amddbarfs0_2rhb0} and 
\ref{amddbarfs0_dens}), the opposite trend of the
mass of particle ($D_s^+$) to be higher than the mass of 
antiparticle ($D_s^+$)
is observed in nuclear matter ($f_s$=0). On the other hand,
the non-zero value of $\langle s^\dagger s \rangle$ 
in the strange hadronic matter leads
to the antiparticle ($D_s^-$) mass to be higher than
that of $D_s^+$ in hyperonic medium. 
Note that $\langle q_i^\dagger q_i \rangle$ for the $D(\bar D)$
mesons corresponds to the number density of the non-strange
quarks ($u$ or $d$), whereas, for $D_s^\pm$, 
$\langle q_i^\dagger q_i \rangle \equiv \langle s^\dagger s\rangle$, 
the number density of the $s$-quark, which is zero 
for nuclear matter.
The mass splitting between the particle and antiparticle 
is thus mainly driven by the term $\langle q_i^\dagger q_i \rangle$, 
which dominates over the contributions
from the dimension 5 operators $\langle {q_i}^\dagger iD_0^2 {q_i}\rangle$
and $\langle {q_i}^\dagger g_s \sigma . G {q_i} \rangle$.
In figure \ref{amdspm_dens}, the mass shifts of the strange-charm mesons
are plotted as functions of the baryon density in units of 
nuclear matter saturation density. With isospin asymmetry,
the mass shifts of both $D_s^\pm$ mesons are observed to be smaller
as compared to the isospin symmetric case for nuclear matter.
The trend is opposite in the presence of strangeness in the medium.
The effect due to isospin asymmetry is observed to be marginal
in strange hadronic matter for T=50 and 100 MeV, whereas, 
at the higher temperature, T=150 MeV, the effect is much larger.

The masses of the $B$ and $\bar B$ mesons are plotted as functions
of the Borel mass in figures \ref{ambbbarfs0_2rhb0} and 
\ref{ambbbarfs3_2rhb0}  for $\rho_B=2\rho_0$
for nuclear matter ($f_s$=0) and strange hadronic matter
(with $f_s$=0.3) respectively, for values of the temperature,
T=50, 100 and 150 MeV. The density dependence of the
mass shifts of $B$ and $\bar B$ mesons are shown in
figures \ref{ambbbarfs0_dens} and \ref{ambbbarfs3_dens} 
for $f_s$=0 and $f_s$=0.3 respectively.
Similar to the cases of $D$ and $\bar D$ mesons
(see figures \ref{amddbarfs0_2rhb0},\ref{amddbarfs3_2rhb0},
\ref{amddbarfs0_dens} and \ref{amddbarfs3_dens}),
the masses of the antiparticles are observed to be larger 
than the masses of the particles in nuclear matter
as well as strange hadronic matter.
For isospin symmetric nuclear (hyperonic) matter, 
the masses of the mesons within the $B(B^+,B^0)$ and 
$\bar B(B^-,\bar {B^0})$
doublets are degenerate. The masses of 
$B$ ($\bar B)$ mesons for T=50, 100
and 150 MeV are obtained as 5384.5 (5349.44), 5378.12 (5343.77) and
5374.27 (5340.35) for $\rho_B=\rho_0$ and 5462.26 (5384), 5453 (5376.87)
and 5446.69 (5372) for $\rho_B=2\rho_0$ in symmetric nuclear matter.
For $\rho_B=\rho_0$, the mass shift of $B$ ($\bar B$) 
in isospin symmetric nuclear matter 
from the vacuum value of 5277.8 MeV is thus 
observed to be 106.7 (71.64), 100.32 (66) and 96.47 (62.55),
corresponding to $B-\bar B$ mass splittings as
35.06, 34.32 and 33.92 respectively.  
These values are similar to the value of the $D-\bar D$
mass splitting of around 32 MeV in isospin symmetric nuclear
matter at $\rho_B=\rho_0$.
The mass shifts (in MeV) from the vacuum values
for the $B$ and $\bar B$ mesons ($\sim$ 107 and 72) 
are observed to be much larger than the values 
of around 75 and 43 for the $\bar D$ and $D$ mesons
in isospin symmetric nuclear matter.
The effect of isospin asymmetry is observed to be larger
for the antiparticles ($B^+$ and $B^0$) as  
compared to the particles ($B^-$ and $\bar {B^0}$).
There is observed to be a drop (rise) in the mass
of $B^+({B^0}$) within the isospin doublet
$(B^+,B^0)$ and in the mass of 
$\bar {B^0}(B^-)$ within the isospin doublet 
$(\bar {B^0},B^-)$, for isospin asymmetric (with $\eta$=0.5)
nuclear and hyperonic matter as compared to the isospin 
symmetric cases. However, the isospin asymmetry effect 
is observed to be marginal for the mass of $\bar {B^0}$.

In fig. \ref{ambs0bar_2rhb0}, the Borel curves 
for the strange-bottom mesons ($B_s^0$ and $\bar {B_s}^0)$
are shown for density $\rho_B=2 \rho_0$ and
Fig. \ref{ambs0bar_dens} shows the density dependence of
the mass shifts of these mesons for the nuclear and 
the strange hadronic matter. 
The mass of the antiparticle (${B_s^0}$) is observed 
to have lower mass as compared to the mass of the 
particle ($\bar {B_s^0}$) in nuclear matter ($f_s$=0) 
for which $\langle q_i ^\dagger q_i\rangle 
\equiv \langle s^\dagger s \rangle$ is zero, whereas,
the opposite trend is observed when there is non-zero strangeness
in the medium. The lower (higher) mass of the
particle, $\bar {B_s^0}$ as compared to the antiparticle,
$B_s^0$ in the nuclear (hyperonic) matter shows that
the term $\langle s^\dagger s\rangle$ dominates
over the other two terms in the odd part of the 
spectral function, which is responsible for the splitting
of the masses between the particle and its antiparticle. 
The behaviour of the mass splitting at finite densities
is similar to the case of strange-charm mesons 
(as shown in Figs. \ref{amdspm_2rhb0} and \ref{amdspm_dens}).
Similar to the strange-charm sector, 
where the isospin asymmetry leads to a drop (rise) 
of the masses of both $D_s^+$ and $D_s^-$ mesons
in nuclear (hyperonic) matter,
the masses of both $\bar {B_s^0}$ and ${B_s^0}$ mesons
are also observed to have a downward (upward) shift
in nuclear (hyperonic) medium (see figures
 \ref{ambs0bar_2rhb0} and \ref{ambs0bar_dens}).
 However, the mass  
modifications due to isospin asymmetry are observed
to be marginal for T=50 and 100 MeV in hyperonic matter.

For the sake of comparison, we show the density dependence
of the mass shifts of the mesons studied in the present work, 
in the hot nuclear and hyperonic matter in figures
\ref{am_charm_fs0_dens}  and \ref{am_charm_fs3_dens} 
for the charm sector and in figures 
\ref{am_bottom_fs0_dens}  and \ref{am_bottom_fs3_dens} 
for the bottom sector. 
The masses are plotted for the isospin symmetric
($\eta$=0) as well as for asymmetric matter (with $\eta$=0.5)
for values of the temperature, T=50, 100 and 150 MeV.
The mass shifts of the open heavy flavor
mesons with the non-strange quark (antiquark) constituents
are observed to be larger than the strange-charm (strange-bottom)
mesons. As has already been mentioned, the antiparticles
($D^-$ and $\bar {D^0}$ for the charm sector and
$B^+$ and $B^0$ for the bottom sector) have larger
mass shifts, as compared to the particles
($D^+$ and $D^0$ for the charm sector and
$B^-$ and $\bar {B^0}$ for the bottom sector), both 
in nuclear and hyperonic matter. The same trend of the
antiparticle having a larger mass shift  as compared
to the particle is observed in hyperonic matter, 
for both strange-charm and strange-bottom mesons, as
might be observed in figures 
\ref{am_charm_fs3_dens}  and \ref{am_bottom_fs3_dens},
whereas the opposite trend is observed when the
strangeness fraction in the medium is zero, 
i.e. in nuclear medium
(see figures \ref{am_charm_fs0_dens} and 
\ref{am_bottom_fs0_dens}). This opposite behaviour of
the strange-heavy flavor mesons in nuclear and hyperonic
media, as has already been seen from figures 
\ref{amdspm_2rhb0}, \ref{amdspm_dens},
\ref{ambs0bar_2rhb0} and \ref{ambs0bar_dens},
arises due to the term proportional to the strange 
quark number density in the odd part of the spectral function
for hyperonic matter, which is zero in nuclear matter.
The isospin asymmetry in the medium
leads to a splitting in the masses between the mesons 
of the iso-doublets ($(D^0,D^+)$, $(\bar {D^0},D^-)$ for the
open charm mesons and $(B^-,\bar {B^0})$, $(B^+,B^0)$ for the
open bottom mesons). As may be observed from figures 
\ref{am_charm_fs0_dens} and \ref{am_charm_fs3_dens},
due to isospin asymmetry in the medium (shown for
isospin asymmetry parameter, $\eta$=0.5), there is an upward 
(downward) shift in the mass of the $D^-$ ($\bar {D^0}$)
for the antiparticle doublet, whereas, there is a drop (rise)
in the mass of $D^+$ ($D^0$) for the particle iso-doublet,
in both nuclear and hyperonic matter.  The isospin asymmetry
is observed to lead to a much larger particle--antiparticle mass
difference for the charged sector ($D^+$ and $D^-$),
whereas, there is observed to be a reduction
in the particle--antiparticle mass difference for the neutral
($D^0$ and $\bar {D^0}$) mesons.
%The $D^+$--$D^-$ mass difference  in isospin asymmetric nuclear
%matter is observed to be  .... and ... for $\rho_B=\rho_0$ and  
%$\rho_B=3\rho_0$ in symmetric ($\eta$=0) nuclear matter 
%at T=100 MeV, which is 
%observed to be modified to ... and ... for these densities
%in asymmetric nuclear matter with $\eta$=0.5.
%On the other hand, for the neutral mesons, the mass difference of ...
%and ...  for densities $\rho_0$ and  $3\rho_0$ 
%symmetric nuclear matter at T=100 MeV are observed
%to be reduced to ... and ... in asymmetric nuclear matter
%for these densities. 
Hence, the effects of the isospin asymmetry 
might modify the particle ratios, $D^+/D^-$ and  $D^0/\bar {D^0}$,
as well as, the dilepton yield from the $D\bar D$ in asymmetric
heavy ion collison experiments.
It might be noted here that the results discussed 
as well as shown in the figures in the present work, correspond
to the maximum possible isospin asymmetry ($\eta$=0.5),
which is, however, not realized in the heavy ion collision experiments.
For a smaller value of isospin asymmetry parameter, $\eta$=0.3 
and T=100 MeV, one observes the particle--antiparticle mass difference 
(in MeV) to be modifed to around 
%1903.36846-1942.683683=39.315223
39 and 157
%1912.2384-2069.248=157.0096
for densities $\rho_0$ and $3\rho_0$ respectively, 
%stated above,
for the charged open charm mesons, and,
21 
%1906.968-1928.0521=21.0841
and
83 
%1934.0696-2017.1756=83.106
for the $D^0$ and $\bar {D^0}$ mesons, from the values
of 30
%1905.67128-1935.81445=30.14317
and 121
%1925.46354-2046.4508=120.98726
for the charged as well as neutral open charm mesons 
%and 30
%%1905.671-1935.8144=30.1434
%and 121
%%1925.4635-2046.45086=120.98736
%for the neutral mesons 
in isospin symmetric ($\eta$=0) nuclear matter.
The effect due to isospin asymmetry is thus observed
to be appreciable, especially at high densities. 

The effect of the baryon density is observed to be the dominant 
medium effect for the modifications of the masses of the open 
charm and open bottom mesons considered in the present work. 
There are significant contributions due to the isospin asymmetry 
of the medium, leading to appreciable increase (decrease)
in the particle--antiparticle mass difference in the charged
(neutral) mesons at high densities, for both charm ($D$, $\bar D$) 
and bottom ($B$, $\bar B$) sectors. 
With accessible energy for study of charm mesons 
at the Compressed Baryonic Matter (CBM) Experiment 
at the Facility for Antiproton and Ion Research (FAIR) 
at the future facility of GSI 
\cite{TheCBMPhysicsBook,P_Senger_HIC_FAIR_NICA_Energies},
which will be designed to investigate matter at high densites
and moderate temperatures, 
the medium modifications of the masses of the $D$ and $\bar D$ mesons
could have observable consequences in the particle ratios,
e.g., $D^+/D^-$, $D^0/{\bar {D^0}}$ in asymmetric nuclear collisions
at CBM, FAIR. The medium modifications of the $D$ and $\bar D$ mesons 
can modify the spectral properties of the charmonium states, 
which can subsequently affect the production
of the open and hidden charmed mesons. 
Also, the dilepton yield arising semi-leptonically
from the $D\bar D$ correlated pairs
can be modified in the isospin asymmetric hadronic matter.
The measurement of the invariant dilepton spectra 
by an electron pair spectrometer at the J-PARC 50-GeV 
proton spectrometer is being carried out
to explore the chiral symmetry of QCD at around nuclear 
density through study of the medium modifications 
of the vector mesons produced 
in pA collsions and decaying to $e^+e^-$ pairs
\cite{jparc_16_phi}, with the E16 experiment at J-PARC 
specifically designed to study the mass of the $\phi$ mesons, 
due to its narrow width. 
The $K^*$ and $K_1$ mesons can be 
produced at J-PARC facility using the kaon beam,
and, partial chiral symmetry restoration in the hadronic
medium can lead to reduction in the mass difference 
of these chiral partners. However,
due to hadronic rescattering, there is expected
to be enhancement in the production ratio of 
$K^+/K^*$, which can be a probe for chiral symmetry
resoration \cite{kstrk1_csb_jparc_16}.
In the future, with upgradation of the spectrometer
at the J-PARC facility, the measurement of 
charm hadrons could be possible. The measurements of various 
hadrons can give a better understanding of the medium 
modifications of the hadron masses as well as a better insight 
to chiral symmetry restoration at high densities
\cite{csb_proposal_Jparc}. 

\section{Summary}

In the present work, we have studied the medium modifications
of the masses of the open charm ($D$, $\bar D$, $D_s$ and $\bar D_s$)
and open bottom ($B$, $\bar B$, $B_s$ and $\bar B_s$) 
in isospin asymmetric strange hadronic matter at finite temperature.
These are computed using a QCD sum rule approach retaining 
the QCD operators upto dimension 5 in the operator product
exapnsion. The values of the QCD operators are calculated
within a chiral SU(3) model. The mass splitting of the particles
($D^0$, $D^+$, $D_s^+$, $B^-$, $\bar {B^0}$, $\bar {B_s^0}$)
and their antiparticles 
($\bar {D^0}$, $D^-$, $D_s^-$, $B^+$, ${B^0}$, ${B_s^0}$)
arises due to the odd part of the spectral function.
In nuclear matter, the masses of the particles, $D^+(D^0)$
and $B^-(\bar {B^0})$ mesons 
are observed to be smaller than the masses of the antiparticles, 
$D^-({\bar D}^0)$ and $B^+({B^0})$. 
For the strange-charm as well as strange-bottom mesons, 
the opposite behaviour of the masses of particles ($D_s^+$ and $\bar {B_s^0}$) 
to be higher than the masses of
antiparticles ($D_s^-$ and $B_s^0$) is observed in nuclear matter, 
whereas, in the strange hadronic matter, 
the non-zero value of $\langle q_i^\dagger q_i \rangle \equiv
\langle s^\dagger s \rangle$ leads 
to the masses of the antiparticles ($D_s^-$ and $B_s^0$) to be higher than
that of $D_s^+$ and $\bar {B_s^0}$ in hyperonic medium.
The sign of the difference in the masses
of the particle and antiparticle ($m_+-m_-$) is observed to
be opposite in nuclear and hyperonic matter,
both for the strange-charm as well as strange-bottom mesons, 
In the mass splitting between the particle and antiparticle 
arising due to the odd part of 
the spectral function, the term proportional
to $\langle q_i^\dagger q_i\rangle
\equiv \langle s^\dagger s \rangle$ for the strange-charm 
(strange-bottom) mesons thus dominates over the other 
QCD operators $\langle {q_i}^\dagger iD_0^2 {q_i}\rangle$
and $\langle {q_i}^\dagger g_s \sigma . G {q_i} \rangle$.
The effects from density and isospin asymmetry, particularly,
on the masses of the $D$, $\bar D$ , $B$ and $\bar B$
mesons are observed to be appreciable. These can modify
e.g., the production of the charmonia and open charm mesons,
and, the production ratios $D^+/D^-$, $D^0/\bar {D^0}$  
in asymmetric relativistic heavy ion
collision experiments at Compressed Baryon Matter 
(CBM) Experiment at the future facility at GSI.

\end{document}